\documentclass[sn-mathphys,Numbered]{sn-jnl}

\usepackage{graphicx}%
\usepackage{multirow}%
\usepackage{amsmath,amssymb,amsfonts}%
\usepackage{amsthm}%
\usepackage{mathrsfs}%
\usepackage[title]{appendix}%
\usepackage{xcolor}%
\usepackage{textcomp}%
\usepackage{manyfoot}%
\usepackage{booktabs}%
\usepackage{algorithm}%
\usepackage{algorithmicx}%
\usepackage{algpseudocode}%
\usepackage{listings}%
\usepackage{hyperref}
\usepackage{color}
\usepackage{graphicx}
\usepackage{bm}
\usepackage{amsmath}
\usepackage{cleveref}
\usepackage{enumerate}
\usepackage{upgreek}
\usepackage[subnum]{cases}
\usepackage{dsfont}
\usepackage{bm}
\usepackage{mathtools}

\newcommand{\tr}{ \text{tr} }

\newcommand{\GHZ}{ \text{GHZ} }

\newcommand{\ti}{ \tilde }

\newcommand{\pa}{ \partial }
\newcommand{\hb}{ \hbar }
\newcommand{\si}{ \sigma }
\newcommand{\om}{ \omega }

\newcommand{\ga}{ \gamma }

\newcommand{\la}{ \langle }
\newcommand{\ra}{ \rangle }

\newcommand{\da}{ \dagger }
\newcommand{\Del}{ \Delta }
\newcommand{\lam}{ \lambda }
\newcommand{\Lam}{ \Lambda }
\newcommand{\vep}{ \varepsilon }
\newcommand{\vrho}{ \varrho }

\newcommand{\GMC}{ \text{GMC} }
\newcommand{\GTC}{ \text{GTC} }

\newcommand{\W}{ \text{W} }
\newcommand{\Wb}{ \overline{\text{W}} }
\newcommand{\vN}{ \text{vN} }
\newcommand{\ESD}{ \text{ESD} }
\newcommand{\con}{ \mathcal{C} }

\newcommand{\gGHZ}{ \text{gGHZ} }
\newcommand{\gW}{ \text{gW} }
\newcommand{\hc}{ \text{h.c.} }

\begin{document}

\title{ Tripartite Entanglement dynamics: the influence of intrinsic decoherence and decoherence channels }


\author*[1]{\fnm{S. V.} \sur{Mousavi}}\email{vmousavi@qom.ac.ir}

\affil*[1]{Department of Physics, University of Qom, Ghadir Blvd., Qom 371614-6611, Iran}

\abstract{
This study examines a system of three coupled qubits, focusing on entanglement measures in the presence of decoherence. It utilizes an XXZ Heisenberg chain with an external magnetic field and Dzyaloshinskii-Moriya interaction, considering intrinsic decoherence. The results reveal that only the magnetic field strength affects entanglement, while intrinsic decoherence suppresses it, with stronger decoherence leading to greater suppression. Various decoherence channels are analyzed, showing that the $I$-tangle typically decreases with increased decoherence, except for the generalized W state under phase damping channel, where only one qubit is affected. Interestingly, dark periods of $I$-tangle occur for the GHZ state under non-Markovian dephasing, and while steady-state entanglement disappears in this channel, it remains nonzero when starting from a mixture of GHZ and fully separable states. Additionally, under generalized amplitude damping channel, reduced bipartite states of a W state exhibit entanglement sudden death, while the steady-state $I$-tangle for the spectral decomposed state stays nonzero.
}


\maketitle

\section{Introduction} \label{intro}

One of the distinguishing aspects of quantum mechanics is the quantum entanglement, which as a resource is used to perform certain tasks such as quantum teleportation and dense coding \cite{Wo-PTRSA-1998, HoHoHoHo-RMP-2009}. In this connection, a crucial concern pertains to the identification of entangled states and quantitative criteria for determining magnitude of the entanglement, have been extensively discussed in bipartite systems. 
Quantum concurrence \cite{HiWo-PRL-1997, Wo-PRL-1998}, $I$-concurrence \cite{RuCa-PRA-2003}, negativity \cite{ViWe-PRA-2002}, log negativity \cite{Pl-PRL-2005} and entropy of entanglement \cite{St-book-2015} are few measures of entanglement introduced for different objectives and applications. 
Concurrence for qubits is defined in terms of the spin flip operator, whereas its generalized version, which is defined for higher-dimensional systems, is referred to as $I$-concurrence due to its definition in terms of the universal-inverter superoperator \cite{RuBuCaHiMi-PRA-2001, RuCa-PRA-2003}.
For a joint pure state $ |\psi\ra $ describing a $ d_A \times d_B $ system, $I$-concurrence is related to the purity of reduced states. This notion is extended to mixed sates via the convex roof construction.
The alternative measure has been proposed as a quantifier of entanglement is tangle, which for pure states is merely the squared concurrence, whereas for mixed states being called $I$-tangle, it is the minimum of the weighted sum of squared concurrences of all pure state decompositions of the mixed state.

In multipartite systems, the qualitative definitions of separability and entanglement, as well as the quantification of entanglement, are considerably more extensive than in the bipartite scenario, owing to the more extensive mathematical structures involved. See \cite{LiSh-FR-2024, Dhetal-book-2017} for recent reviews. There are three distinct classes of pure tripartite states: fully separable, bipartite separable and genuine entangled \cite{GuTo-PR-2009}. While fully separable states are the product of three single particle states, bipartite separable states are written as a product state in the bipartite system which  is formed when two of three subsystems grouped together to one party. Finally, a pure state is referred to as genuinely tripartite entangled if it is not separable, fully or bipartite. 
%Greenberger-Horne-Zeilinger ($\GHZ$) state and the so-called W state lie in this category.
%
This last class of the genuinely entangled three-qubit states is itself divided into two subclasses under stochastic local operations and classical communication (SLOCC). The Greenberger-Horne-Zeilinger ($\GHZ$)-class states
$ | \psi_{\GHZ{\text{-c}}} \ra = \cos\varphi~|000\ra + \sin\varphi~|11\ra \otimes ( \cos\varphi_3 |0\ra + \sin\varphi_3 |1\ra ) $
and the so-called W-class states 
$ | \psi_{\W{\text{-c}}} \ra = a |001\ra + b |010\ra + \sqrt{1 - a^2 - b^2} |100\ra $; $a$ and $b$ being real, are representative elements of the two different SLOCC classes \cite{Sz-PRA-2011, AjRu-PRA-2010}. 
The GHZ state, $ \varphi = \pi/4 $ and $ \varphi_3 = \pi/2 $, is maximally entangled in the sense that its one-qubit reduced states are maximally mixed.
On the other hand, its bipartite subsystems are separable having diagonal density matrix. On the contrary, the one-partite subsystems of the $W$ state, $a = b = 1 / \sqrt{3}$, are less mixed than the ones of the GHZ state, but its bipartite subsystems are entangled with concurrence $2/3$ \cite{Sz-PRA-2011}.
A three-qubit mixed state is fully (bipartite) separable if it can be written as a convex combination of fully (bipartite) separable pure states. Otherwise, the state is genuinely entangled \cite{GuSe-NJP-2010}. 
A classification of entanglement in three-qubit systems has been presented based on three-qubit and reduced two-qubit entanglements \cite{SaGa-EPJD-2008}.
Various measures have been proposed for quantification of entanglement in tripartite systems, including residual entanglement \cite{CKW-PRQA-2000}, $\pi$-tangle \cite{OuFa-PRA-2007}, $I$-tangle \cite{Os-PRA-2005}, genuine tripartite concurrence \cite{MaChChSpGaHu-PRA-2011}, and concurrence fill \cite{XiEb-PRL-2021, GeLiCh-PRA-2023}. 
The last measure has been employed to quantify the tripartite entanglement in three-flavour neutrino oscillations \cite{LiLiSoWaYe-EPJC-2022}.

The constraints for sharing quantum correlations between different parties are characterized by monogamy relations, which are distinct properties of quantum correlations in multipartite systems \cite{Dhetal-book-2017}.
A bipartite quantum correlation measure $ Q $ applied to a quantum state $ \rho_{ABC} $ is termed as monogamous if the relation
%
%\begin{eqnarray} \label{eq: monogam}
$ Q_{A|BC} \geq Q_{AB} + Q_{AC} $
%\end{eqnarray}
%
holds; $ Q_{AB} $ and $ Q_{AC} $ being respectively correlations of the reduced states $ \rho_{AB} $ and $ \rho_{AC} $ while $ Q_{A|BC} $ quantifies correlation contained in the sate $ \rho_{ABC} $ considered in the $ A | BC $ bipartite split, taking $ BC $ as a single object \cite{WaLiLiSoWa-EPJC-2023}.
The first monogamy relation, known as CKW inequality, was proposed in \cite{CKW-PRQA-2000} taking the squared concurrence as the measure of quantum correlation. 
%When the state of the three-qubit system is pure, the concurrence of the pair $ A | BC $ i.e., $ C_{A|BC} $ is $ 2 \sqrt{\det(\rho_A)} $; $\rho_A$ being the reduced state describing the qubit $A$.
%
Using the concept of interaction information, the necessary and sufficient conditions for the measure to adhere to monogamy have been identified for arbitrary pure and mixed quantum states. It has been found that while three-qubit generalized GHZ (gGHZ) states follow monogamy, generalized W (gW) states do not \cite{PrPaSeSe-PRA-2012}.
A general monogamy inequality in a multiqubit mixed state has been derived for  the squared entanglement of formation \cite{BaXuWa-PRL-2014}. The general monogamy and polygamy relationships for different powers of entanglement measures and assisted entanglement measures have been examined \cite{XiZhLi-QIP-2023}.
Multipartite entanglement of a one-dimensional anisotropic spin$-1/2$ XXZ spin chain with the Dzyaloshinskii-Moriya (DM) interaction have been considered \cite{ChWuXu-QIP-2017}. 
The role of DM interaction and how it affects the quantum entanglement and leading to the entanglement sudden death (ESD) has been investigated in the literature \cite{ShPa-QIP-2014, ShPa-QIP-2015, ShPa-QIP-2016, ShPa-QIP-2016_2}. 
It has been found that the DM interaction affects the entanglement distribution and enhance the proportion of multipartite entanglement in the entanglement structure. Furthermore, an analytical relation has been given for $ \uptau(\rho_{ABC}) = E_f^2(\rho_{B|AC}) - E_f^2(\rho_{BA}) - E_f^2(\rho_{BC}) $, $E_f$ being the entanglement of formation. It has bee illustrated that for a given strength of DM interaction, $ \uptau(\rho_{ABC}) $ decreases with anisotropicity $\Delta$. 
For a recent review and some generalized version of monogamy inequalities see \cite{ZoYiSoCa-FP-2022}.

Quantum systems are not isolated from their environment. The existence of the environment causes decoherence, which affects quantum properties such as coherence or linear superposition, as well as the entanglement.
As stated by Le Bellac, ``The Enemy Number One of a quantum computer is decoherence, the interaction of qubits with the environment which blurs the delicate linear superpositions." \cite{Be-book-2006}. Therefore, it is an important task to study how the environment affects the entanglement.  
Two commonly used complementary approaches to describe an open quantum system are the Lindbladian master equations and quantum operations or quantum channels \cite{NiCh-book-2000}. In the first formalism, an evolution equation, the so-called master equation, which contains non-unitary terms, is obtained for the reduced density operator describing the system of interest. 
For a recent illustration, see \cite{Mo-EPJP-2024} to find out how, within this formalism, different quantum correlations including entanglement, are affected by decoherence.
The second complementary approach involves Kraus operators acting on the initial density matrix mapping it to another state.  
Quantum properties under decoherence channels have been widely explored in the literature \cite{BaKiSh-JPA-2005, YaLiYiGuHa-PLA-2011, DuWaYe-SR-2017, SlDaLa-QIP-2019, LiXiHaFaHeGu-SR-2023, HuWaSu-PRA-2011, ChBhPa-EPJP-2022, GuYaTiWaZe-QIP-2019, ChFuFa-IJTP-2014, GuZeCh-LPL-2019}.

A distinct perspective pertains to the modification of the conventional quantum mechanical evolution equation, the von Neumann equation, wherein certain terms have been incorporated into it. These additional terms induces decoherence, the intrinsic one. One such method has been proposed by Milburn \cite{Mi-PRA-1991}. 
The Milburn equation has been utilized in the literature for both continuous variable systems \cite{MoMi-JPA-2022, MoMi-EPJP-2024} and finite dimensional systems with discrete degrees of freedom. The effects of intrinsic decoherence on various quantum correlations and quantum features have been extensively studied in bipartite systems using diverse models \cite{EsKhMaDa-OPE-2022, MoKhHaRaTaPo-RP-2022, Haetal-APB-2022, Daetal-OQE-2023, MuCh-QIP-2021, AiHaNa-QIP-2021, QuChMa-IJTP-2022, QuMa-PS-2024, MoRaEl-Ent-2022, MoEl-SR-2022, NaMuCh-PA-2022, ChMaa-APB-2023, HoSeNoLi-SC-2024}. However, only a few studies have been devoted to quantum correlations in tripartite systems with intrinsic decoherence. 
The influence of the intrinsic decoherence on the residual entanglement and tripartite uncertainty bound in an XXZ Heisnberg chain \cite{WaIbAmSa-EPJD-2024}, as well as on the entanglement dynamics in the Tavis-Cummings model \cite{HoWaMa-EPJD-2012} has been examined. 
Furthermore, considering a Heisenberg XY chain with three qubits in an external magnetic field, the effect of intrinsic decoherence on the Bell-nonlocality sudden death has been studied \cite{LiShZo-PLA-2010}.

In the present manuscript, our objective is to investigate dynamics of the entanglement in a tripartite system consisting of three qubits. Different measures including residual entanglement, $I$-tangle, concurrence fill and genuine tripartite concurrence will be used to quantify the entanglement. Decoherence will be taken into account by decoherence channels and the intrinsic decoherence in the Milburn framework. We consider only those states with those channels where the entanglement measures can be computed analytically. These states include pure states and second rank density operators.

This work is organized as follows. 
Different entanglement measures based on concurrence are introduced in the tripartite systems in Section \ref{sec: Entan}.
Section \ref{sec: en_ev_Sch} deals with evolution of entanglement in the framework of the Schr\"odinger equation. Here, evolution of the gW state is studied; and squared concurrences and concurrence fill will be considered. 
Section \ref{sec: en_ev_Mil} is devoted to the examination of the impact of intrinsic decoherence on entanglement within the framework of the Milburn equation.
The effect of decoherence channels on tripartite entanglement measures will be considered in section \ref{sec: channels}.
Section \ref{sec: results} devotes to results and discussions.
Finally, summary and conclusions will be presented in Section \ref{sec: conclusion}. 

%\section{Quantum entaglement: concurrence and negativity} \label{sec: QC}

\section{An overview of entanglement measures in a tripartite system composed of three qubits} \label{sec: Entan}

An entanglement measure $E$ fulfils the following conditions (i) $ E(\rho) = 0 $ when the state $\rho$ is fully separable or biseparable (ii) $E$ is non-increasing under local operations and classical communications (LOCC), i.e., it is an entanglement monotone (iii) The function $E$ is invariant under local unitary operations, i.e., $ E(\rho) = E( U_A \otimes U_B \otimes U_C ~ \rho ~ U_A^{\da} \otimes U_B^{\da} \otimes U_C^{\da}  ) $ for any local unitary operators $ U_A $, $U_B$ and $U_C$. 
One may consider SLOCC as a stronger condition for entanglement ordering \cite{LiSh-FR-2024}.
Tripartite quantum entanglement can be quantified using different measures. One is the {\it residual entanglement} or {\it three-tangle} defined as
\begin{eqnarray} \label{eq: residual_entang}
\uptau(|\psi \ra) &=& \con^2_{A|BC}(|\psi \ra) - \con^2_{AB} - \con^2_{AC}
\end{eqnarray}
for the pure state $ |\psi \ra $ where $ \con^2_{ij} $ refers to the squared concurrence of the reduced state $ \rho_{ij} = \tr_k(|\psi\ra\la \psi|) $ and $ \con^2_{A|BC} $ denotes the entanglement between subsystems $A$ and $BC$, as a single object, and is known as {\it $I$-tangle}. %In this way, three-tangle implies the amount of entanglement, quantified by the squared concurrence, between parties $A$ and $BC$, as a single object, that is not contained in the entanglements of $A$ with $B$ and $C$ separately \cite{CKW-PRQA-2000}. 
It is noteworthy that $ \uptau(|\psi \ra) $ remains invariant under any qubit chosen as the focus qubit.  
Via the convex roof construction, the residual entanglement is generalized to the case of mixed states: $ \uptau( \rho ) = \min_{ \{ p_i, |\psi_i\ra \} } \sum_i p_i ~  \mathcal{\uptau}( | \psi_i \ra ) $, where minimization is performed over all possible pure state decompositions of the mixed state.   
%
%The three-tangle has been termed the {\it monogamy score} of the squared concurrence in {\color{blue}the} recent literature \cite{Dhetal-book-2017}. 
The three-tangle has recently been termed as the {\it monogamy score} of the squared concurrence in the literature \cite{Dhetal-book-2017}. 
Generally, based on the monogamy relation fulfilled by a given quantum correlation, the corresponding monogamy score is defined. The monogamy score of the squared negativity is referred to as {\it $\pi$-tangle}, which can be easily computed as it does not rely on the convex roof construction \cite{OuFa-PRA-2007}. Contrary to three-tangle, the $\pi$-tangle depends on the focus qubit \cite{OuFa-PRA-2007}. The so-called {\it three-$\pi$} defined as the average of three $\pi$-tangles corresponding to three different focus qubits is invariant under permutations of the qubits.
For the pure state $ | \psi \ra $, one obtains
\begin{eqnarray} \label{eq: C_A|BC_pure}
\con_{A|BC}( | \psi \ra ) &=& \sqrt{ 2 [ 1 - \tr( \rho_A^2 ) ] } =  2 \sqrt{ \det( \rho_A ) },
\end{eqnarray}
$ \rho_A $ being the reduced state describing the qubit $A$. % i.e.,  $ \rho_A = \tr_{BC}( | \psi \ra \la \psi | ) $. 
Again, the prescription of convex roof construction is used to evaluate the entanglement for different bi-partitions corresponding to a mixed state, $ \con^2_{A|BC}( \rho ) = \min_{ \{ p_i, |\psi_i\ra \} } \sum_i p_i ~  \con^2_{A|BC}( | \psi_i \ra ) $. 
However, when the mixed state is a second rank matrix i.e., has only two nonzero eigenvalues, $ \con^2_{A|BC} $ can be computed analytically through \cite{Os-PRA-2005}
\begin{eqnarray} \label{eq: C2_A|BC_rank2}
\con_{A|BC}^2(\rho) &=& \tr( \rho \ti{\rho} ) + 2 m_{\min} [ 1 - \tr( \rho^2 ) ] ,
\end{eqnarray}
where 
\begin{eqnarray}
\ti{\rho} &=& \mathds{1}_2 \otimes \mathds{1}_4 - \rho_A \otimes \mathds{1}_4 - \mathds{1}_2 \otimes \rho_{BC} + \rho ,
\end{eqnarray}
and $ m_{\min} $ is the minimum eigenvalue of the real symmetric $ 3 \times 3 $ matrix $ M $ whose elements given in the appendix \ref{app: M-matrix} by Eqs. \eqref{eq: M11}, \eqref{eq: M12}, \eqref{eq: M13}, \eqref{eq: M22}, \eqref{eq: M23} and \eqref{eq: M33}.
Note that in Eq. \eqref{eq: C2_A|BC_rank2}, the second term of the right hand side includes the linear entropy of the state, $ S_L(\rho) = 1- \tr(\rho^2) $.

Another concept in multipartite systems is the {\it genuine entanglement}. The {\it genuine tripartite concurrence} (GTC) of a pure state $ |\psi \ra $ describing a system composed of three qubits is defined as \cite{MaChChSpGaHu-PRA-2011}
\begin{eqnarray} \label{eq: C-gtc}
\con_{\GTC}(|\psi \ra) &=& \min \{ \con_{A|BC}, \con_{B|AC}, \con_{C|AB} \},
\end{eqnarray}
%
%%where as mentioned above $ \con_{i|jk} = \sqrt{ 2 [ 1 - \tr_i(\tr_{jk} (\rho))^2 ] } $ is the entanglement between qubit $``i"$ and the remaining two qubits taken together as an ``other" single party, the {\it one-to-other bipartite entanglement}.
which is generalized to mixed states via the convex roof construction \cite{WuZe-EPJC-2022}.
%
%\begin{eqnarray} \label{eq: gen_con}
%\mathcal{C}_{\GTC}(\rho) &=& \min_{ \{ p_i, |\psi_i \ra \} } \sum_i p_i \mathcal{C}_{\GMC}( |\psi_i \ra ).
%\end{eqnarray}
%
%due to the relation \eqref{eq: C2_A|BC_rank2}, this quantity can be computed analytically for a rank-two mixed state. 
For the X-shaped state
\begin{eqnarray} \label{eq: tripar_Xstate}
\rho_X &=& 
\left(
\begin{array}{cccccccc}
 n_1 & 0 & 0 & 0 & 0 & 0 & 0 & c_1 \\
 0 & n_2 & 0 & 0 & 0 & 0 & c_2 & 0 \\
 0 & 0 & n_3 & 0 & 0 & c_3 & 0 & 0 \\
 0 & 0 & 0 & n_4 & c_4 & 0 & 0 & 0 \\
 0 & 0 & 0 & c_4^* & m_4 & 0 & 0 & 0 \\
 0 & 0 & c_3^* & 0 & 0 & m_3 & 0 & 0 \\
 0 & c_2^* & 0 & 0 & 0 & 0 & m_2 & 0 \\
 c_1^* & 0 & 0 & 0 & 0 & 0 & 0 & m_1 \\
\end{array}
\right),
\end{eqnarray}
it has the analytic form
\begin{eqnarray}  \label{eq: gen_con_X}
\con_{\GMC}(\rho_X) &=& 2 \max \{ 0, |c_i| - \nu_i \}, \quad i = 1, \cdots, 4,
\end{eqnarray}
where $ \nu_i = \sum_{j \neq i }^4 \sqrt{ n_j m_j } $.

By considering the concurrence triangle, the so-called {\it concurrence fill}
\begin{eqnarray} \label{eq: entang-fill}
\mathcal{F}(|\psi \ra) &=& \left[ \frac{16}{3} \mathcal{Q} ( \mathcal{Q} - \con^2_{A|BC} ) ( \mathcal{Q} - \con^2_{B|AC} ) ( \mathcal{Q} - \con^2_{C|AB} ) \right]^{1/4}
\end{eqnarray}
being proportional to the square root of the area of the concurrence triangle, has been recently proposed \cite{XiEb-PRL-2021} as another genuine tripartite entanglement measure for the pure state $ | \psi \ra$. Here,
\begin{eqnarray} \label{eq: half-perimeter}
\mathcal{Q} &=& \frac{1}{2} ( \con^2_{A|BC} + \con^2_{B|AC} + \con^2_{C|AB} )
\end{eqnarray}
is the half of the perimeter of the concurrence triangle.
The perimeter $ 2 \mathcal{Q} $ has been called the global entanglement, although it may be a feasible measure, but not a genuine one in the sense that it can be nonvanishing for certain biseparable states \cite{GeLiCh-PRA-2023}.
Since concurrence fill \eqref{eq: entang-fill} depends on the lengths of all legs of the concurrence triangle, whereas GTC \eqref{eq: C-gtc} only depends on the shortest edge (indeed its squre root), concurrence fill contains more information than GTC \cite{XiEb-PRL-2021}.   
By using \eqref{eq: residual_entang} in \eqref{eq: entang-fill}, one obtains \cite{GeLiCh-PRA-2023}
\begin{eqnarray} \label{eq: confil}
\mathcal{F}(|\psi \ra) &=& \left[ \big( \uptau + \frac{2}{3}(\con^2_{AB}+\con^2_{AC}+\con^2_{BC}) \big) ( \uptau + 2\con^2_{AB} ) ( \uptau + 2\con^2_{AC} ) ( \uptau + 2\con^2_{BC} ) \right]^{1/4} .
\end{eqnarray}
%
%where $ \con^2_{AB} = \con^2(\rho_{AB}) $ with $ \rho_{AB} = \tr_C(|\psi\ra\la \psi|) $ being the reduced state. 
%Therefore, when all bipartite reduced states are disentangled, the concurrence fill reduces to the residual entanglement.
%
Via the convex roof construction, the concurrence fill is generalized to the case of mixed states \cite{XiEb-PRL-2021}.
%: $ \mathcal{F}_{ABC}( \rho ) = \min_{ \{ p_i, |\psi_i\ra \} } \sum_i p_i ~  \mathcal{F}_{ABC}( | \psi_i \ra ) $, where minimization is performed over all possible pure state decompositions of the mixed state. 

\section{Entanglement evolution in the the Schr\"odinger framework} \label{sec: en_ev_Sch}

Our system is taken an XXZ Heisenberg chain composed of three qubits being in an external homogeneous magnetic field with the DM interaction.
DM interaction is an antisymmetric exchange interaction that originates from the spin orbit coupling. In contrast to the conventional Heisenberg exchange, which typically leads to ferromagnetic or antiferromagnetic spin configurations, DM interaction favors canted or helical-like spin structures \cite{CaLi-SSR-2023, YaLiCu-NRP-2023}. 
Taking all interactions in the $z$ direction then the Hamiltonian of such a system reads \cite{Daetal-OQE-2023}
\begin{eqnarray} \label{eq: Ham}
H &=& J \sum_{i=1}^{3} \bigg\{ \si^{(i)}_x \si^{(i+1)}_x + \si^{(i)}_y \si^{(i+1)}_y + \Del ~ \si^{(i)}_z \si^{(i+1)}_z \bigg\} 
\nonumber \\
& + &
D \sum_{i=1}^{3} ( \si^{(i)}_x \si^{(i+1)}_y - \si^{(i)}_y \si^{(i+1)}_x ) + B \sum_{i=1}^{3} \si^{(i)}_z ,
\end{eqnarray}
where $ \si^{(i)}_x $ represents the $x$ component of the $i^{\mbox{th}}$ Pauli matrix, and coupling between qubits has been denoted by $J$, $ \Del $ being the anisotropy parameter and $ D $ displays the strength of $z$ component of DM interaction. 
Note that one assumes the periodic boundary condition $ \si^{(i+3)} = \si^{(i)} $. 

If the system is initially described by the pure state $ | \psi(0) \ra $, its evolution is given by $ | \psi(t) \ra = e^{ - i H t /\hb} | \psi(0) \ra $. 
The gGHZ state 
\begin{eqnarray} \label{eq: geGHZ-state}
| \gGHZ \ra &=& a | 000 \ra + \sqrt{1-a^2} | 111 \ra ,
\end{eqnarray}
then evolves to
\begin{eqnarray} \label{eq: geGHZ-Sch}
| \gGHZ \ra_t &=& a ~ e^{-i 3 t(B + J \Del)} | 000 \ra + \sqrt{1-a^2} ~ e^{i 3 t(B - J \Del)} | 111 \ra ,
\end{eqnarray}
where we have put $ \hb = 1 $. As one sees, this state is independent of the strength of DM interaction $D$. The corresponding density operator is then given by
\begin{eqnarray} \label{eq: geGHZ-rho-Sch}
\rho(t) &=& | \gGHZ \ra_t ~ _t\la \gGHZ | \nonumber \\
&=& a^2  | 000 \ra \la 000 | + a \sqrt{1-a^2} ~ e^{-i 6 B t } | 000 \ra \la 111 | + a \sqrt{1-a^2} ~ e^{i 6 B t } | 111 \ra \la 000 | 
\nonumber \\
&~&+ (1-a^2) | 111 \ra \la 111 | ,
\end{eqnarray}
which depends only on the magnetic field strength $B$. From this, one obtains
\begin{eqnarray} \label{eq: geGHZ-bi-Sch}
\rho_{AB} &=& \rho_{AC} = \rho_{BC} = a^2 | 00 \ra \la 00 | + ( 1 - a^2 ) | 11 \ra \la 11 |
\end{eqnarray}
for the bipartite reduced states and finally 
\begin{eqnarray} \label{eq: geGHZ-one-Sch}
\rho_A &=& \rho_B = \rho_C =  a^2 | 0\ra \la 0 | + ( 1 - a^2 ) | 1 \ra \la 1 |
\end{eqnarray}
for the one particle reduced states. These relations demonstrate that entanglement measures for the gGHZ state have no dynamics under the Schr\"odinger evolution equation. Explicitly, $ \mathcal{C}_{AB} = \mathcal{C}_{AC} = \mathcal{C}_{BC} = 0 $, $ \mathcal{C}_{A|BC} = \mathcal{C}_{B|AC} = \mathcal{C}_{C|AB} = 2 \sqrt{a^2(1-a^2)} $, $\uptau = 4 a^2(1-a^2) $ and $ \mathcal{C}_{\GTC} = 2 \sqrt{a^2(1-a^2)} $. From this, one sees that nonzero concurrences are maximum for $ a = 1 / \sqrt{2} $ i.e., for GHZ state.
As Eq. \eqref{eq: confil} shows, in order for concurrence fill to be different from the residual entanglement, at least one of the reduced bipartite states must be entangled. The gGHZ state \eqref{eq: geGHZ-state} does not fulfil this requirement. 

In the next step, we examine entanglement properties of the gW state
\begin{eqnarray}  \label{eq: geW-state} 
| \gW \ra &=& a | 001 \ra + b | 010 \ra + \sqrt{1-a^2-b^2} | 100 \ra .
\end{eqnarray}
One sees all reduced states $ \rho_{AB}(t) $, $ \rho_{AC}(t) $ and $ \rho_{BC}(t) $ are X-shaped. Thus the corresponding concurrence of the state $ \rho_{AB}(t) $ is given by
\begin{eqnarray} \label{eq: concur-X}
\mathcal{C}_{AB}(t) &=& 2 \max \bigg\{0, |(\rho_{AB}(t))_{14}| - \sqrt{(\rho_{AB}(t))_{22} (\rho_{AB}(t))_{33}}, |(\rho_{AB}(t))_{32}| 
\nonumber \\
&~&- \sqrt{(\rho_{AB}(t))_{11} (\rho_{AB}(t))_{44}}  \bigg\}.
\end{eqnarray}
Since $ (\rho_{AB}(t))_{14} = (\rho_{AB}(t))_{44} = 0 $, the concurrence recasts
\begin{eqnarray} \label{eq: concur-AB-geW-Sch}
\mathcal{C}_{AB}(t) &=& 2 |(\rho_{AB}(t))_{32}|.
\end{eqnarray}
The same is true for other reduced bipartite states.
Because the relationship for the elements of the density operator is lengthy, we do not bring them here and analyze only some properties.
Although the evolved state depends on all parameters appearing in the Hamiltonian, one sees $ (\rho_{ij}(t))_{32} $ and thus concurrences are independent of both the anisotropy parameter $ \Del $ and the field strength $B$. 
Trigonometric functions appearing in the matrix elements $ (\rho_{AB}(t))_{32} $, $ (\rho_{AC}(t))_{32} $ and $ (\rho_{BC}(t))_{32} $ have five different frequencies of oscillation, 
\begin{equation}
\om_1 = 6|J|, \quad \om_2 = 2 \sqrt{3} | D |, \quad \om_3 = 4 \sqrt{3} | D |, \quad \om_4 = | 2 \sqrt{3} D - 6 J |, \quad \om_5 = | 2 \sqrt{3} D + 6 J | .
\end{equation}
In the case that these frequencies are commensurate--that is if
\begin{eqnarray}
\frac{\om_1}{n_1} = \frac{\om_2}{n_2} = \frac{\om_3}{n_3} = \frac{\om_4}{n_4} = \frac{\om_5}{n_5} ,
\end{eqnarray}
where $n_1, n_2, n_3, n_4$ and  $n_5$ are integers, the reduced bipartite concurrences are periodic.
% with period 
%
%\begin{eqnarray} \label{eq: red-cocur-geW-Sch-period}
%T_{ \mathcal{C}_{ij} } &=& 2\pi \times \min \left\{ \frac{n_1}{\om_1}, \frac{n_2}{\om_2}, \frac{n_3}{\om_3} \right\}
%\end{eqnarray}
%
From this analysis one sees that in the absence of DM interaction i.e., for $D=0$, the period of bipartite concurrences is $ \pi /(3|J|) $. For the special values $ D = 1/\sqrt{3} $ and $ J = \pm 1 $, the period is $ \pi /2 $. 
Trigonometric functions appearing in the determinant of one-particle states have nine different frequencies of oscillation $ \{  4 \sqrt{3} |D|, 8 \sqrt{3} |D|, 4 | \sqrt{3} D - 3 J |, 12 |J|, 6| \sqrt{3} D - J |, 6 | \sqrt{3} D + J |, 4| \sqrt{3} D + 3 J |,  2| \sqrt{3} D - 3 J |,  2| \sqrt{3} D + 3 J | \} $. Again, if these frequencies are commensurate then the one-to-other concurrences $ \mathcal{C}_{A|BC} $, $ \mathcal{C}_{B|AC} $ and $ \mathcal{C}_{C|AB} $ given by \eqref{eq: C_A|BC_pure} are periodic.
%Thus, from Eq. \eqref{eq: C-gtc} one sees the genuine tripartite concurrence $ \mathcal{C}_{\GTC} $ has the same period while Eq. \eqref{eq: confil} show that the period of concurrence fill is half of the period of GTC. 
For the special case $ D = 0 $ this period is $ \pi /(3|J|) $ while for $ J = \pm 1 $ this period is $ \pi /2 $ for $ D = 1/\sqrt{3} $ and $ \pi/6 $ for $ D = \sqrt{3} $.
Furthermore, calculations indicate that the residual entanglement is identical to zero,
\begin{eqnarray}
\uptau(| gW \ra_t) &=& 0 ,
\end{eqnarray}
$ | gW \ra_t $ being the evolution of the state \eqref{eq: geW-state} under the Schr\"odinger equation.

\section{ Entanglement dynamic in the framework of Milburn equation } \label{sec: en_ev_Mil}

Postulating that the system, on sufficiently short-term steps, does not evolve continuously under unitary evolution, Milburn \cite{Mi-PRA-1991} obtained the following master equation for the evolution of the system:
\begin{eqnarray} \label{eq: Milburn0}
\frac{\pa \rho}{\pa t} &=& \ga \left( e^{-i H / \hb \ga} \rho ~ e^{i H / \hb \ga} - \rho \right) ,
\end{eqnarray}
where $H$ is the Hamiltonian of the system and $\ga$ is the well-known intrinsic decoherence parameter (the inverse of the minimum time step in the universe responsible for a stochastic sequence of identical unitary transformations at sufficiently short time steps). 
Expansion of Eq. (\ref{eq: Milburn0}) up to first order in $\ga^{-1}$ reads  
\begin{eqnarray} \label{eq: Milburn}
\frac{\pa \rho}{\pa t} &=& - \frac{i}{\hb} [H, \rho] - \frac{1}{2\hb^2 \ga} [H, [H, \rho] ] ,
\end{eqnarray}
which simplifies to the familiar von Neumann equation at the lowest order. 
In the energy eigenbasis, the Milburn equation \eqref{eq: Milburn0} reads
\begin{eqnarray} \label{eq: Mil0_rhot_nnp}
\frac{\pa}{\pa t}\varrho_{n n'}(t) &=& \ga \left( e^{-i (E_n - E_{n'}) / \hb \ga}  -1  \right) \varrho_{nn'},
\end{eqnarray}
where $ \{ E_n \} $ is the set of eigenvalues of the Hamiltonian \eqref{eq: Ham};
\begin{numcases}~ 
E_1 = -3 ( B - J \Delta ) , \label{eq: E1}\\
E_2 = - B - 2 \sqrt{3} D - J ( \Del + 2 ) , \\
E_3 = B - 2 \sqrt{3} D - J ( \Del + 2 ) , \\
E_4 = - B + 2 \sqrt{3} D - J ( \Del + 2 ) , \\
E_5 = B + 2 \sqrt{3} D - J ( \Del + 2 ) , \\
E_6 = -B - J ( \Del - 4 ) , \\
E_7 = B - J ( \Del - 4 ) , \\
E_8 = 3 ( B + J \Del ) , \label{eq: E8}
\end{numcases}
and $ \varrho_{n n'}(t) = \la E_n | \rho(t) | E_{n'} \ra $ represents the density matrix elements in the energy representation, where
$ \{ | E_n \ra \} $ denotes the associated energy eigenfunctions,
\begin{eqnarray} 
H | E_n \ra &=& E_n | E_n \ra  .
\end{eqnarray}
Solution of Eq. \eqref{eq: Mil0_rhot_nnp} is 
\begin{eqnarray} \label{eq: rhot_nnp}
\varrho_{nn'}(t) &=& \varrho_{nn'}(0) ~ f_{n n'}(t) ,  
\end{eqnarray}
where we have introduced
\begin{eqnarray} \label{eq: f_ij}
f_{n n'}(t) &=& \exp\left[\ga \left( e^{-i (E_n - E_{n'}) / \hb \ga}  -1  \right) t \right],
\end{eqnarray}
which in the limit $ \ga^{-1} \to 0 $ recasts
\begin{eqnarray} \label{eq: f_ij-approx}
f_{n n'}(t) &\approx& \exp\left[ - i \frac{(E_n - E_{n'})t}{\hb} - \frac{ (E_n - E_{n'})^2 t }{2\hb^2 \ga} \right].
\end{eqnarray}
Thus, for the density operator one has that
\begin{eqnarray} \label{eq: rhot_abs}
\rho(t) &=& \sum_n \sum_{n'} \varrho_{nn'}(t) |E_n\ra \la E_{n'}| \\
&=&  \sum_n \sum_{n'} \varrho_{nn'}(0) ~ f_{n n'}(t)~ |E_n\ra \la E_{n'}| .
\end{eqnarray}

The transformation unitary matrix $ U $ that relates the computational basis, $ \{ | 000 \ra, | 001 \ra, | 010 \ra, | 100 \ra, | 011 \ra, | 101 \ra, | 110 \ra, | 111 \ra \} $, to the energy basis, $ \{ | E_1 \ra, | E_2 \ra, | E_3 \ra, | E_4 \ra, | E_5 \ra, | E_6 \ra, | E_7 \ra, | E_8 \ra \} $, is characterized by the elements $ U_{mn} = \la a_m | E_n \ra $, where $ |a_m\ra $ represents elements of the computational basis. It has the following explicit form:
\begin{equation} \label{eq: U-matrix}
U = \left(
\begin{array}{cccccccc}
 0 & 0 & 0 & 0 & 0 & 0 & 0 & 1 \\
 0 & 0 & \frac{1}{6} \left(-\sqrt{3}+3 i\right) & 0 & \frac{1}{6} \left(-\sqrt{3}-3 i\right) & 0 & \frac{1}{\sqrt{3}} & 0 \\
 0 & 0 & \frac{1}{6} \left(-\sqrt{3}-3 i\right) & 0 & \frac{1}{6} \left(-\sqrt{3}+3 i\right) & 0 & \frac{1}{\sqrt{3}} & 0 \\
 0 & 0 & \frac{1}{\sqrt{3}} & 0 & \frac{1}{\sqrt{3}} & 0 & \frac{1}{\sqrt{3}} & 0 \\
 0 & -\frac{1}{2 \sqrt{3}}+\frac{i}{2} & 0 & \frac{1}{6} \left(-\sqrt{3}-3 i\right) & 0 & \frac{1}{\sqrt{3}} & 0 & 0 \\
 0 & \frac{1}{6} \left(-\sqrt{3}-3 i\right) & 0 & \frac{1}{6} \left(-\sqrt{3}+3 i\right) & 0 & \frac{1}{\sqrt{3}} & 0 & 0 \\
 0 & \frac{1}{\sqrt{3}} & 0 & \frac{1}{\sqrt{3}} & 0 & \frac{1}{\sqrt{3}} & 0 & 0 \\
 1 & 0 & 0 & 0 & 0 & 0 & 0 & 0 \\
\end{array}
\right) .
\end{equation}

Taking the initial state as the gGHZ state \eqref{eq: geGHZ-state}
%
%\begin{eqnarray} \label{eq: GHZ0}
%|\Psi \ra &=& a | 000 \ra + \sqrt{1 - a^2} | 111 \ra
%\end{eqnarray}
%
the evolved density operator under the Milburn equation \eqref{eq: Milburn0} in the energy eigenbasis is given by
\begin{eqnarray} \label{eq: rho-ghz-en}
\varrho(t) &=& (1 - a^2) |E_1 \ra \la E_1 | + f_{18}(t) a \sqrt{1 - a^2} |E_1 \ra \la E_8 |
\nonumber \\
&+& f_{81}(t) a \sqrt{1 - a^2} |E_8 \ra \la E_1 | + a^2  |E_8 \ra \la E_8 | ,
\end{eqnarray}
where $ f_{nn'}(t) $ is given by \eqref{eq: f_ij}.
Then the evolved density operator in the computational basis becomes
\begin{eqnarray} \label{eq: rhot}
\rho(t) = U \varrho(t) U^{\dag} &=& a^2 | 000 \ra \la 000 | + f_{81}(t)  a \sqrt{1 - a^2} | 000 \ra \la 111 | \nonumber \\
&+& f_{18}(t)  a \sqrt{1 - a^2} | 111 \ra \la 000 | + ( 1 - a^2 ) | 111 \ra \la 111 |.
\end{eqnarray}
The reduced state describing the qubit $A$ is then
\begin{eqnarray} \label{eq: rhot_A}
\rho_A(t) = \tr_{BC}(\rho(t)) = a^2 | 0 \ra \la 0 | + ( 1 - a^2 ) | 1 \ra \la 1 |.
\end{eqnarray}
The state \eqref{eq: rhot} is a second rank matrix with only two non-zero eigenvalues
\begin{eqnarray} \label{eq: rhot_eigenval}
\lam_{1,2} &=& \frac{1}{2} \left\{  1 \pm \sqrt{ 1 - 4a^2(1-a^2)(1-|f_{18}(t) |^2) } \right\} .
\end{eqnarray}
%
%with the corresponding eigenvectors
%
%\begin{eqnarray} \label{eq: rhot_eigenvec}
%
%\end{eqnarray}
%
In such cases the $I$-tangle $ C_{A|BC}^2(t) $ can be calculated analytically by using Eq. \eqref{eq: C2_A|BC_rank2}.
%
%\begin{eqnarray} \label{eq: C2_A|BC_rhot_}
%C_{A|BC}^2(t) &=& \tr( \rho(t) \ti{\rho}(t) ) + 2 m_{\min}(t) [ 1 - \tr( \rho^2(t) ) ]
%\end{eqnarray}
%
%where 
%
%\begin{eqnarray}
%\ti{\rho}(t) &=& \mathds{1}_2 \otimes \mathds{1}_4 - \rho_A(t) \otimes \mathds{1}_4 - \mathds{1}_2 \otimes \rho_{BC}(t) + \rho(t)
%\end{eqnarray}
%
%and $ m_{\min}(t) $ is the minimum eigenvalue of a real symmetric $ 3 \times 3 $ matrix $ M $ given in \cite{Os-PRA-2005} 
For the state \eqref{eq: rhot}, elements of the matrix $M$ whose minimum eigenvalue becomes important are given by \eqref{eq: M11_rhot}, \eqref{eq: M12_rhot}, \eqref{eq: M13_rhot}, \eqref{eq: M22_rhot}, \eqref{eq: M23_rhot} and \eqref{eq: M33_rhot}. 

Finally, by use of Eq. \eqref{eq: mmin_rhot} in \eqref{eq: C2_A|BC_rank2} one obtains
\begin{eqnarray} \label{eq: C2_A|BC_rhot}
C_{A|BC}^2(t) &=& 4 a^2 (1-a^2) | f_{18}(t) |^2 
= 4 a^2 (1-a^2) \exp \left[ - 4 \ga t \sin^2 \left( \frac{3 B}{\ga} \right) \right]
\end{eqnarray}
for the evolved gGHZ state \eqref{eq: rhot} where in the second equality we have used \eqref{eq: E1}, \eqref{eq: E8} and \eqref{eq: f_ij} to obtain
\begin{eqnarray} \label{eq: f_18}
f_{18}(t) &=& \exp\left[\ga \left( e^{ 6 i B / \ga}  -1  \right) t \right].
\end{eqnarray}
Eq. \eqref{eq: C2_A|BC_rhot} shows that the only relevant parameter of the Hamiltonian is the strength of the applied magnetic field. 
Note that in the standard quantum theory described by the von Neumann equation (or the Schr\"odinger equation in the case of pure states), one obtains $ |f_{n n'}(t)| = 1 $. Thus, from \eqref{eq: C2_A|BC_rhot} one obtains
\begin{eqnarray} \label{eq: C2_A|BC_rhot_vN}
C_{A|BC}^2\bigg|_{\vN} &=& 4 a^2 (1-a^2) = 4 \det(\rho_A(t)) ,
\end{eqnarray}
where in the second equality we have used \eqref{eq: rhot_A}. %This result was obtained in the previous section.
%Within the standard quantum mechanics the initial pure state, here the GHZ state, remains pure under the von Neumann evolution equation. It has been already argued that the concurrence between parties $A$ and $BC$ is just $ 2 \sqrt{\det (\rho_A)} $ for pure states \cite{CKW-PRQA-2000}. 
Note that the standard quantum result \eqref{eq: C2_A|BC_rhot_vN} is constant, independent of the Hamiltonian parameters and depend only on the parameter $a$ appearing in the initial state \eqref{eq: geGHZ-state}.  
As a function of $ a $, the squared concurrence in the von Neumann regime, $ C_{A|BC}^2\bigg|_{\vN} $, takes the maximum value $1$ for $ a = 1 / \sqrt{2} $. For this value of $a$, the squared concurrence in the Milburn framework recasts 
\begin{eqnarray} \label{eq: C2_A|BC_rhot_a0}
C_{A|BC}^2(t) \bigg|_{a = 1 / \sqrt{2} } &=& \exp \left[ - 4 \ga t \sin^2 \left( \frac{3 B}{\ga} \right) \right] ,
\end{eqnarray}
where for a given $t$ and $\ga$ is an oscillatory function of $B$ with period $ \ga \pi/3 $.
Eq. \eqref{eq: gen_con_X} yields
\begin{eqnarray} \label{eq: gen_con_rhoGHZg}
C_{\GTC}(t) &=& 2 a \sqrt{1-a^2} ~ |f_{18}(t)| 
= 2 a \sqrt{1-a^2} \exp \left[ - 2 \ga t \sin^2 \left( \frac{3 B}{\ga} \right) \right] ,
\end{eqnarray}
%
%for the genuine concurrence of the {\color{green}gGHZ} state under the intrinsic decoherence in the Milburn framework 
where in the second equality we have used \eqref{eq: f_18}. 
%Comparison with \eqref{eq: C2_A|BC_rhot} reveals that $ C_{\GTC}(t) = C_{A|BC}(t) $ for $a=1/\sqrt{2}$. 

%\subsection{A mixture of GHZ and fully separable states under intrinsic decoherence}

We now consider the effect of the intrinsic decoherence on the mixed state
\begin{eqnarray} \label{eq: rho0_GHZ_000_111}
\rho_0 = (1-w_1-w_2) |\GHZ \ra \la \GHZ | + w_1 |000\ra \la 000| + w_2 |111\ra \la 111| , 
\end{eqnarray}
where $|\GHZ \ra$ is given by \eqref{eq: geGHZ-state} for $a = 1/\sqrt{2}$ i.e.,
\begin{eqnarray} \label{eq: GHZ-state}
| \GHZ \ra &=& \frac{1}{\sqrt{2}} (|000\ra + |111\ra ). 
\end{eqnarray}
Note that the state \eqref{eq: rho0_GHZ_000_111} is fully separable for $ w_1 + w_2 = 1 $, while it remains in the entanglement regime for $ w_1 + w_2 < 1 $, as in the case of the GHZ state.
Evolution of the state \eqref{eq: rho0_GHZ_000_111} under the Milburn equation \eqref{eq: Milburn0} yields 
\begin{eqnarray} \label{eq: rhot-GHZ_000_111}
\rho(t) &=& \frac{1+w_1-w_2}{2} | 000 \ra \la 000 | + f_{81}(t)  \frac{1-w_1-w_2}{2} | 000 \ra \la 111 | \nonumber \\
&+& f_{18}(t) \frac{1-w_1-w_2}{2} | 111 \ra \la 000 | + \frac{1-w_1+w_2}{2} | 111 \ra \la 111 |.
\end{eqnarray}
Using $m_{\min}=-\frac{1}{2}$ for the minimum eigenvalue of the $M$ matrix in \eqref{eq: C2_A|BC_rank2} is then yields
\begin{eqnarray} 
C_{A|BC}^2(t) &=& (1-w_1-w_2)^2 \exp \left[ - 4 \ga t \sin^2 \left( \frac{3 B}{\ga} \right) \right] 
\label{eq: C2_A|BC_GHZ_000_111} \\ 
&=&
(1-w_1-w_2)^2 C_{A|BC}^2(t) \bigg|_{\GHZ} , \label{eq: mixed_pure}
\end{eqnarray}
%
%for the squared one-to-other concurrence 
where in the second equality we have used \eqref{eq: C2_A|BC_rhot_a0}.
Eq. \eqref{eq: mixed_pure} shows that $I$-tangle of the mixed state \eqref{eq: rho0_GHZ_000_111} under the Milburn equation is proportional to the corresponding quantity for the pure GHZ state. However, it is reduced by a factor of  $ (1 - w_1 - w_2)^2 $, which is the square of the weight factor of the GHZ state in the initial mixed state.

Owing to the X-shape of the density matrix \eqref{eq: rhot-GHZ_000_111}, from  Eq. \eqref{eq: gen_con_X} one obtains
\begin{eqnarray} \label{eq: GMC_GHZ_000_111}
\con_{\GMC}(t) &=& (1-w_1-w_2) |f_{18}(t)| = 
(1-w_1-w_2) \exp \left[ - 2 \ga t \sin^2 \left( \frac{3 B}{\ga} \right) \right].
\end{eqnarray}
%
%for the genuine tripartite entanglement. 
From Eqs. \eqref{eq: C2_A|BC_GHZ_000_111} and \eqref{eq: GMC_GHZ_000_111} one  sees that $ \con_{\GMC}(t) = C_{A|BC}(t) $. Furthermore, the entanglement content is maximal for $ w_1 = w_2 = 0 $ i.e., when the system is described by the GHZ state.

At the end of this section, it is worth mentioning that, as stated in the introduction, the Milburn model has been employed in both bipartite and tripartite systems to explore the effects of intrinsic decoherence on nonclassical correlations \cite{MoKhHaRaTaPo-RP-2022} and nonlocality \cite{LiShZo-PLA-2010}. The first paper examines a Heisenberg XYZ chain consisting of two qubits subjected to an external magnetic field and Kaplan-Shekhtman-Entin-Wohlman-Aharony (KSEA) interaction, focusing on how intrinsic decoherence impacts nonclassical correlations. The authors employ log negativity as a measure for entanglement and found that the magnetic field plays a significant role in generating and sustaining nonclassical correlations. Moreover, they discovered that the KSEA interaction can affect the creation, preservation, and revival of these correlations, whereas intrinsic decoherence significantly weakens nonclassical correlations. 
Ref. \cite{LiShZo-PLA-2010} explores the phenomenon of Bell nonlocality sudden death (BNSD) in a three-qubit Heisenberg XY chain influenced by an external magnetic field, framed within the Milburn context. This study evaluates Bell nonlocality by analyzing the degree of violation of multipartite Bell inequalities. It investigates two categories of nonlocal correlations: tripartite correlations linked to the Svetlichny inequality and those pertaining to the WWZB (Werner-Wolf and \.Zukowski-Brukner) inequalities. The results for a system initially characterized by the GHZ state \eqref{eq: GHZ-state} indicate that violations of the Bell inequality can be completely eliminated within a specific timescale, which is notably influenced by both intrinsic decoherence and the external magnetic field.  
In contrast, our focus is on the dynamics of entanglement within this framework, specifically by examining the genuine tripartite concurrence as well as the one-to-other concurrences.

\section{Effects of decoherence channels on entanglement} \label{sec: channels}

A quantum channel is a completely positive trace-preserving quantum map that maps density matrices to density matrices and has a representation in terms of Kraus operators $ \{ K_m \} $ where $ \sum_m K_m^{\dag} K_m = \mathds{1} $, $ \mathds{1} $ being the identity operator. 
When interacting with the environment, various scenarios can be considered, including situations where all qubits are coupled to the environment or asymmetric damping scenarios where only a subset of the qubits are affected by the noise \cite{Ca-QIP-2013}. Kraus operators $ K_m $ are tensor products of Kraus operators $ A_i $ corresponding to single-qubit channels. In our three-qubit system when only the first qubit is coupled to the environment then the channel acts as 
\begin{eqnarray} \label{eq: channel-I}
\rho \to \rho' = \sum_{i}^n (A_i \otimes \mathds{1} \otimes \mathds{1} ) \rho (A_i^{\dag} \otimes \mathds{1} \otimes \mathds{1} ) ,
\end{eqnarray}
$n$ being the number of Kraus operators. 
We refer to such a channel as ``name-I", where ``name" implies the channel's name and ``I" implies that only the first qubit is coupled to the environment. Likewise, ``name-III" refers to a channel where the environment affects only the third qubit.
In general where all qubits are coupled to the environment, the state changes as \cite{LiXiHaFaHeGu-SR-2023}
\begin{eqnarray} \label{eq: st-chan-gen}
\rho \to \rho' = \sum_{ijk} (A_i \otimes A_j \otimes A_k ) \rho ( A_i^{\dag} \otimes A_j^{\dag} \otimes A_k^{\dag}) .
\end{eqnarray}

In this section, we will consider effect of different channels on different initial tripartite states. 
Our main mission is to find an analytical relation for the squared concurrence between subsystems $A$ and $BC$.
% as we are interested in the monogamy of the squared concurrence. 
In this way, we will consider the evolution of initial states with a rank one or two under those channels that maintain the rank of the state below or equal to two.
We are particularly interested in the phase damping channel (PDC), non-Markovian dephasing channel, and amplitude damping channel (ADC). Under the two former ones, the gGHZ state is mapped onto a second rank matrix while the same is true for a superposition of W and $\Wb$ states under ADC.% {\color{red} with the later being the obverse of former}. 
The other channel which we are interested in is the generalized amplitude damping channel (GADC) although it maps the initial state to a matrix with a rank greater than two. For this channel, we study concurrence of reduced sates discovering ESD. Furthermore, as an upper limit for $ C^2_{A|BC} $, we examine spectral decomposed state under this channel.

De Oliveira has already realized that gGHZ (gW) state under local independent PDC (ADC) is mapped onto a second rank matrix and thus the squared concurrences can be computed analytically \cite{Ol-PRA-2009}. He additionally realized that this is not the case for GHZ (W) under ADC (PDC).
But, we have noticed that GHZ and W states are mapped onto second rank matrices under respectively ADC-I and  PDC-I given by \eqref{eq: channel-I}.

%if only one of the qubits, which we take it the first one i.e., the party $A$, interacts with the environment the generalized GHZ and W states are mapped onto rank-2 matrices;
%
%\begin{eqnarray} \label{eq: channel-I}
%\rho \to \rho' = \sum_{m=0}^1 (A_m \otimes \mathds{1} \otimes \mathds{1} ) \rho (A_m^{\dag} \otimes \mathds{1} \otimes \mathds{1} )
%\end{eqnarray}
%   

\subsection{W state under PDC-I}

PDC describes the loss of coherence while keeping the energy intact and is represented by the Kraus operators
\begin{eqnarray} \label{eq: PDC}
A_0 = |0 \ra \la 0 | + \sqrt{1-d} |1 \ra \la 1 |, \qquad 
A_1 =  \sqrt{d} |1 \ra \la 1 |,
\end{eqnarray}
$d$ being the decoherence strength \cite{LiXiHaFaHeGu-SR-2023}.
The W state 
\begin{eqnarray} \label{eq: W-state}
| W \ra = \frac{1}{\sqrt{3}} ( | 001 \ra + | 010 \ra + | 100 \ra )
\end{eqnarray}
under PDC-I is mapped onto
\begin{eqnarray} \label{eq: rhopW}
\rho'_{\W} &=& 
\left(
\begin{array}{cccccccc}
 0 & 0 & 0 & 0 & 0 & 0 & 0 & 0 \\
 0 & \frac{1}{3} & \frac{1}{3} & 0 & \frac{\sqrt{1-d}}{3} & 0 & 0 & 0 \\
 0 & \frac{1}{3} & \frac{1}{3} & 0 & \frac{\sqrt{1-d}}{3} & 0 & 0 & 0 \\
 0 & 0 & 0 & 0 & 0 & 0 & 0 & 0 \\
 0 & \frac{\sqrt{1-d}}{3} & \frac{\sqrt{1-d}}{3} & 0 & \frac{1}{3} & 0 & 0 & 0 \\
 0 & 0 & 0 & 0 & 0 & 0 & 0 & 0 \\
 0 & 0 & 0 & 0 & 0 & 0 & 0 & 0 \\
 0 & 0 & 0 & 0 & 0 & 0 & 0 & 0 \\
\end{array}
\right).
\end{eqnarray}
which is a second rank density matrix. For the minimum eigenvalue of the corresponding $M$-matrix appearing in \eqref{eq: C2_A|BC_rank2} we obtain
\begin{equation} \label{eq: mmin-gW-PDCI}
m_{\min} = \frac{ \sqrt{ (32 d - 41) (32 d - 33) } - 1 }{8 (8 d-9)} ,
\end{equation}
and thus,
\begin{eqnarray} \label{eq: C2_A|BC-rhopW}
\mathcal{C}^2_{A|BC}(\rho'_{\W}) &=& \frac{d \left(\sqrt{(32 d-41) (32 d-33)}-1\right)}{9 (8 d-9)}-\frac{4 (d-2)}{9}.
\end{eqnarray}
%
%which is minimum at $ d \approx 0.9280 $ with the minimal value $0.14203$. 
For other one-to-other squared concurrences we obtain
\begin{equation} \label{eq: C2_B|AC-rhopW}
\mathcal{C}^2_{B|AC}(\rho'_{\W}) = \mathcal{C}^2_{C|AB}(\rho'_{\W}) = 
\frac{d \left(\sqrt{1296 d^2-2792 d+1497}+4 d-5\right)}{18 (8 d-9)}-\frac{2}{9} \left(2 d+\sqrt{1-d}-5\right) .
\end{equation}

The reduced states $ \rho'_{AB} $ and $ \rho'_{AC} $ are the same being X-shaped with squared concurrences
\begin{eqnarray} \label{eq: C_AB-rhopW}
\mathcal{C}^2(\rho'_{AB}) &=& \mathcal{C}^2(\rho'_{AC}) = 4 \frac{1-d}{9} ,
\end{eqnarray}
while squared concurrence of the reduced state $ \rho'_{BC} $ is constant
\begin{eqnarray} \label{eq: C_BC-rhopW}
\mathcal{C}^2(\rho'_{BC}) &=& \frac{ 4 }{9} ,
\end{eqnarray}
a not an unexpected result as the environment only affects the first qubit.
From these relations and based on the relation $ \uptau = \min [\mathcal{C}^2_{A|BC}] - \mathcal{C}^2_{AB} - \mathcal{C}^2_{AC} $ which has been mentioned in \cite{OuFa-PRA-2007}, it's tempting to conclude that the residual entanglement depends on the focus qubit. But, as stated in the introduction, the residual entanglement for mixed states should be computed via the convex roof construction. Thus, the mentioned relation of \cite{OuFa-PRA-2007} seems to be incorrect.

\subsection{A mixture of W and $|000\ra $ states under ADC}

Consider the mixture 
\begin{eqnarray} \label{eq: W_000_mix}
\rho &=& (1-w ) |W\ra \la W| + w |000\ra \la 000 |
\end{eqnarray}
of the entangled state $|W\ra$ given by \eqref{eq: W-state} and the fully separable state $|000\ra $. This mixed state being a second rank matrix, remains second rank under ADC. The single-qubit Kraus operators corresponding to ADC are \cite{NiCh-book-2000}
\begin{eqnarray} \label{eq: ADCI}
A_0 = |0 \ra \la 0 | + \sqrt{1-d} |1 \ra \la 1 |, \qquad 
A_1 =  \sqrt{d} |0 \ra \la 1 | ,
\end{eqnarray}
where the decoherence strength $d$ represents the probability of decay from the upper level to the lower one and it is related to the damping rate $\ga$ via the relation $ d = 1 - e^{-2\ga t} $ \cite{NiCh-book-2000, HuWaSu-PRA-2011}. This quantum channel describes the probability of losing an excitation to the environment and thus affects
both the coherence and energy within a system. 
Then by using \eqref{eq: ADCI} in \eqref{eq: st-chan-gen} one obtains 
\begin{eqnarray} \label{eq: rhop_W_000}
\rho' &=& 
%\left(
%\begin{array}{cccccccc}
 %d+(1-d)w & 0 & 0 & 0 & 0 & 0 & 0 & 0 \\
 %0 & \frac{(1-d)(1-w)}{3} & \frac{(1-d)(1-w)}{3} & 0 & \frac{(1-d)(1-w)}{3} & 0 & 0 & 0 \\
% 0 & \frac{(1-d)(1-w)}{3} & \frac{(1-d)(1-w)}{3} & 0 & \frac{(1-d)(1-w)}{3} & 0 & 0 & 0 \\
% 0 & 0 & 0 & 0 & 0 & 0 & 0 & 0 \\
% 0 & \frac{(1-d)(1-w)}{3} & \frac{(1-d)(1-w)}{3} & 0 &\frac{(1-d)(1-w)}{3} & 0 & 0 & 0 \\
% 0 & 0 & 0 & 0 & 0 & 0 & 0 & 0 \\
% 0 & 0 & 0 & 0 & 0 & 0 & 0 & 0 \\
% 0 & 0 & 0 & 0 & 0 & 0 & 0 & 0 \\
%\end{array}
%\right)
[ d+(1-d)w ] | 000 \ra \la 000 | + \frac{(1-d)(1-w)}{3} \bigg( | 001 \ra \la 001 | + | 001 \ra \la 010 | + | 001 \ra \la 100 | 
\nonumber \\
& + & | 010 \ra \la 010 | + | 010 \ra \la 100 | + | 100 \ra \la 100 | + \text{h.c.} \bigg)
\end{eqnarray}
for the evolved state where ``h.c." implies ``Hermitian conjugate". 
The linear entropy of the state is given by
\begin{eqnarray} \label{eq: linent_W_000}
S_L(\rho') &=& 1 - \tr(\rho'^2) = 2d(1-d) + (4 d^2 - 6 d + 2) w - 2 (1-d)^2 w^2 ,
\end{eqnarray}
which is a measure of the impurity or mixedness of the state. As indicated by this equation, the state is ultimately pure when $d=1$, regardless of the value of the weight factor $w$.

As stated the rank of this density matrix is two. Computations reveal that the minimum eigenvalue of the corresponding $M$-matrix appearing in \eqref{eq: C2_A|BC_rank2} is zero. Thus, for the squared one-two-other concurrence one has that
\begin{eqnarray} \label{eq: C2_A|BC-rhopWmix}
\mathcal{C}^2_{A|BC}(\rho') &=& \frac{8}{9} (1-d)^2 (1-w)^2.
\end{eqnarray}
All reduced states have the same form
\begin{eqnarray}
\rho'_{AB} &=& \rho'_{BC} = \rho'_{AC} =
\left(
\begin{array}{cccc}
 \frac{1}{3} (2 d (1-w)+2 w+1) & 0 & 0 & 0 \\
 \\
 0 & \frac{(1-d)(1-w)}{3} & \frac{(1-d)(1-w)}{3} & 0 \\
 \\
 0 & \frac{(1-d)(1-w)}{3} & \frac{(1-d)(1-w)}{3} & 0 \\
 \\
 0 & 0 & 0 & 0 \\
\end{array}
\right) ,
\end{eqnarray}
with the same concurrence
\begin{eqnarray} \label{eq: reducedConcur-rhopWmix}
\mathcal{C}(\rho'_{AB}) &=& \mathcal{C}(\rho'_{BC}) = \mathcal{C}(\rho'_{AC}) = \frac{ 4 }{9} (1-d)^2 (1-w)^2 .
\end{eqnarray}

\subsection{Generalized GHZ state under ADC-I and ADC-III}

In effect of ADC-I given by \eqref{eq: ADCI},  from \eqref{eq: channel-I} one sees that the gGHZ state \eqref{eq: geGHZ-state} is mapped onto
\begin{eqnarray} \label{eq: rhopGHZ}
\rho'_{\gGHZ} &=& 
\left(
\begin{array}{cccccccc}
 a^2 & 0 & 0 & 0 & 0 & 0 & 0 & a \sqrt{1-a^2} \sqrt{1-d} \\
 0 & 0 & 0 & 0 & 0 & 0 & 0 & 0 \\
 0 & 0 & 0 & 0 & 0 & 0 & 0 & 0 \\
 0 & 0 & 0 & d(1-a^2) & 0 & 0 & 0 & 0 \\
 0 & 0 & 0 & 0 & 0 & 0 & 0 & 0 \\
 0 & 0 & 0 & 0 & 0 & 0 & 0 & 0 \\
 0 & 0 & 0 & 0 & 0 & 0 & 0 & 0 \\
 a \sqrt{1-a^2} \sqrt{1-d} & 0 & 0 & 0 & 0 & 0 & 0 & \left(a^2-1\right) (d-1) \\
\end{array}
\right) ,
\end{eqnarray}
which is a second rank matrix. 
Thus, the squared concurrence can be calculated analytically.
For the state \eqref{eq: rhopGHZ},  $ m_{\min} = 0 $ and thus the squared coherence between parties $A$ and $BC$ obtained from \eqref{eq: C2_A|BC_rank2} is
\begin{eqnarray} \label{eq: C2_A|BC-rhopGHZ}
\mathcal{C}^2_{A|BC}(\rho'_{\gGHZ}) &=& 4 a^2 ( 1- a^2 ) ( 1-d ) .
\end{eqnarray}
As this equation shows, in a given $d$ the squared concurrence $ \mathcal{C}^2_{A|BC}(\rho'_{\GHZ}) $ is maximum for $ a = 1/\sqrt{2} $.
One can easily check that the concurrence of reduced states $ \rho'_{AB} $ and $ \rho'_{AC} $ are both zero and thus monogamy score or three-tangle of GHZ state under ADC-I is just the squared coherence between parties $A$ and $BC$. 
As can be seen from \eqref{eq: gen_con_X} and \eqref{eq: C2_A|BC-rhopGHZ}, $ \mathcal{C}_{\GTC} = \sqrt{ \mathcal{C}^2_{A|BC} } $.

In the next step, we consider the situation where only the third qubit i.e., qubit $C$, is in interaction with the environment. Then, one has that 
\begin{eqnarray} \label{eq: channel-III}
\rho \to \rho'' = \sum_{m=0}^1 ( \mathds{1} \otimes \mathds{1} \otimes A_m ) \rho ( \mathds{1} \otimes \mathds{1} \otimes A_m^{\dag} ).
\end{eqnarray}
In this case, we obtain
\begin{eqnarray} \label{eq: C2_A|BC-rhopGHZ-3}
\mathcal{C}^2_{A|BC}(\rho''_{\gGHZ}) &=& 2 a^2 (1 - a^2) (2 - d ),
\end{eqnarray}
which is not zero at $ d = 1 $ for $ a \neq 1 $. For the genuine tripartite concurrence one gets
\begin{eqnarray} \label{eq: genC-rhopGHZ-3}
\mathcal{C}_{\GTC} &=& 2 a \sqrt{(1 - a^2)(1-d)} ,
\end{eqnarray}
which is different than $ \sqrt{ \mathcal{C}^2_{A|BC}(\rho''_{\GHZ}) } $.

\subsection{A mixture of GHZ and $|000\ra $ states under PDC}

In this section, we study the effect of PDC \eqref{eq: PDC} on the mixture
\begin{eqnarray} \label{eq: GHZ_000_mix}
\rho &=& (1-w) | \GHZ \ra \la \GHZ | + w |000\ra \la 000 |
\end{eqnarray}
of the entangled GHZ state \eqref{eq: GHZ-state} and the fully separable state $|000\ra $. In the situation where the environment affects all qubits, the evolved state is given by
\begin{eqnarray} \label{eq: rhop_GHZ_000}
\rho' &=& \left( \frac{1 + w}{2} \right) | 000 \ra \la 000 | + \frac{(1-d)^{3/2}(1-w)}{2}
 \left( | 000 \ra \la 111 | + \text{h.c.} \right) 
\nonumber\\
&+& 
  \frac{1-w}{2} | 111 \ra \la 111 |.
\end{eqnarray}
The linear entropy, a measure of mixedness, of this state is then given by
\begin{eqnarray} \label{eq: LinEnt_GHZ_000_mixed}
S_L(\rho') &=& \frac{1}{2} \left(1 -(1-d)^3\right)+(1-d)^3 w - \frac{1}{2} \left( 1 + (1-d)^3\right) w^2 .
\end{eqnarray}
%
%Here, the final state is generally mixed $ S_L(\rho')\big|_{d=1} = (1-w^2)/2 $. Furthermore, 
The impurity of the final state is higher than that of the initial state. 
%This is in opposite to the situation of W state under ADC.
% 
From Eq. \eqref{eq: gen_con_X} one obtains
\begin{eqnarray} \label{eq: GHZ_000_mix_GMC}
\con_{\GMC}(\rho') &=& (1-d)^{3/2} (1-w)
\end{eqnarray}
for the genuine tripartite entanglement. %All reduced states are the same and independent of the decoherence rate $d$, having the X-shape $  \frac{1+w}{2}| 000 \ra \la 000 | + \frac{1-w}{2} | 111 \ra \la 111 |  $ are separable.
All reduced states are identical and do not depend on the decoherence rate $d$. They exhibit an X-shaped form, given by $  \dfrac{1+w}{2}| 000 \ra \la 000 | + \dfrac{1-w}{2} | 111 \ra \la 111 | $, and are separable.

The minimum eigenvalue of matrix $M$ appearing in Eq. \eqref{eq: C2_A|BC_rank2} is lengthy and will not be given here. For the power series expansion of the squared one-to-other concurrence one then obtains
\begin{eqnarray} \label{eq: C2_A|BC-rhopGHZmix}
\mathcal{C}^2_{A|BC}(\rho') &=& (1 - d)^3 (1- w)^2 + O\left(w^4\right) ,
\end{eqnarray}
which is valid for $ w \ll 1 $. This equation and Eq. \eqref{eq: GHZ_000_mix_GMC} show that when a mixture of GHZ and $|000\ra$ states, wherein the weight factor of the later is extremely small, undergoes PDC, the entanglement content,  in comparison to that of GHZ state, decreases with a factor $1-w$.

\subsection{Generalized GHZ state under the non-Markovian dephasing channel } 

The non-Markovian dephasing channel \cite{Ho-PA-2020} which is rooted in the open system dynamics modeling random telegraph noise \cite{SSB-PRA-2018, PaPaBaSr-QIP-2021} is described by the Kraus operators
\begin{eqnarray} \label{eq: nonMarkov-ch}
A_0 &=& \sqrt{ \frac{ 1 + \Lam(t) }{2} } \mathds{1} ,  \qquad 
A_1 = \sqrt{ \frac{ 1 - \Lam(t) }{2} } \si_z, 
\end{eqnarray}
% 
%$ \mathds{1} $ being the identity $ 2 \times 2 $ matrix, $ \si_z $ the Pauli spin matrix 
where
\begin{eqnarray} \label{eq: Lambda}
\Lam(t) &=& e^{-\nu} \left( \cos(\mu ~ \nu) + \frac{1}{\mu} \sin(\mu ~ \nu)\right), 
\quad  \mu = \sqrt{(4 b \tau)^2 -1}, \quad \nu = \frac{t}{2\tau} ,
\end{eqnarray}
with $b$ as the magnitude of random field and $ \tau^{-1} $ the fluctuation rate. In the Markovian regime, $\mu$ is an imaginary number i.e., $ (4 b \tau)^2 < 1 $  \cite{Pa-QIP-2016}.
The gGHZ state under this channel is mapped onto
\begin{eqnarray} \label{eq: gGHZ_nonMarkov}
\rho'_{\gGHZ}(t)\bigg|_{\text{GADC}}&=& a^2 | 000 \ra \la 000 | + a \sqrt{1 - a^2} \Lam^3(t) ~ | 000 \ra \la 111 | \nonumber 
\\
&+&  a \sqrt{1 - a^2} \Lam^3(t) ~ | 111 \ra \la 000 | + ( 1 - a^2 ) | 111 \ra \la 111 |,
\end{eqnarray}
where we have assumed all qubits are affected by the channel.
Here, the reduced states describing bipartite systems $AB$, $AC$ and $BC$ are all the same: $ a^2 | 00 \ra \la 00 | + (1-a^2) | 11 \ra \la 11 | $. Thus, bipartite concurrences are all zero. %, $ \mathcal{C}(\rho'_{AB}) = \mathcal{C}(\rho'_{AC}) = \mathcal{C}(\rho'_{BC}) =0 $. 
From \eqref{eq: C2_A|BC_rank2} one obtains
\begin{eqnarray} \label{eq: C2_A|BC-nonMarkov_gGHZ}
\mathcal{C}^2_{A|BC}(t) &=& 2 a^2 (1-a^2) [ 1+\Lam^6(t) + 2 m_{\min}(t) (1-\Lam^6(t)) ],
\end{eqnarray}
which for GHZ state with $ m_{\min}(t) = - 1/2 $ recasts
\begin{eqnarray} \label{eq: C2_A|BC-nonMarkov}
\mathcal{C}^2_{A|BC}(t) \bigg|_{\GHZ}&=& \Lam^6(t).
\end{eqnarray}
The monogamy score of the squared concurrence is then $ \mathcal{C}^2_{A|BC}(t) $.
Eq. \eqref{eq: gen_con_X} yields
\begin{eqnarray} \label{eq: gen_con_rhoGHZg_nonMarkov}
\mathcal{C}_{\GTC}(t) &=& 2 a \sqrt{1-a^2} ~ |\Lam^3(t)| .
\end{eqnarray}
%
%for the genuine concurrence of the {\color{green}gGHZ} state under the non-Markovian dephasing channel.

%\subsection{A mixture of GHZ and $  |000\ra $ states under the non-Markovian dephasing channel } 

For the evolution of the mixture \eqref{eq: GHZ_000_mix} under the dephasing channel \eqref{eq: nonMarkov-ch} one obtains
\begin{eqnarray} \label{eq: GHZ000_nonMarkov}
\rho' &=& \frac{1+w}{2} | 000 \ra \la 000 | + \frac{1-w}{2} \Lam^3(t) ( | 000 \ra \la 111 | + \hc ) +  \frac{1-w}{2} | 111 \ra \la 111 |,
\end{eqnarray}
and from which
\begin{eqnarray} 
\mathcal{C}^2_{A|BC}(t) &=& \frac{1}{2} (1-w) \bigg( 1 + w + (1-w)\Lam^6(t) \bigg) \label{eq: C2A_BC_nonMar_GHZ_mix}
\nonumber \\
&+& (1-w) \bigg( 1 + w - (1-w)\Lam^6(t) \bigg) m_{\min}(t)  , \\
\mathcal{C}_{\GTC}(t) &=& (1-w) ~ |\Lam^3(t)| ,
\end{eqnarray}
%
%respectively for one-to-other concurrence and the tripartite genuine concurrence 
where $m_{\min}(t)$ is given by
\begin{eqnarray}
m_{\min}(t) &=& \frac{w^2-\sqrt{40 (1-w)^2 w^2 \Lam^6(t) + 16 (1-w)^4 \Lam^{12}(t) + 9 w^4}}{8 \left((1-w)^2 \Lam^6(t) + w^2\right)}.
\end{eqnarray}

\subsection{Generalized $\W\Wb$ state under GADC} \label{sec: model}

The generalized amplitude damping channel (GADC) is described by the four Kraus operators \cite{NiCh-book-2000}
\begin{eqnarray} \label{eq: GADC-channel}
A_0 &=& \sqrt{p} ( |0 \ra \la 0 | + \sqrt{1-d} |1 \ra \la 1 | ), \qquad \qquad 
A_1 = \sqrt{p} \sqrt{d} |0 \ra \la 1 |, \nonumber \\
A_2 &=& \sqrt{1 - p} ( \sqrt{1-d} |0 \ra \la 0 | + |1 \ra \la 1 | ), \qquad 
A_3 = \sqrt{1-p} \sqrt{d} |1 \ra \la 0 | ,
\end{eqnarray}
where $p$ is the probability of finding the sysytem in the ground state when the system is in thermal equiblirium \cite{TeDeFrSaOlBo-PhTRS-2012}.
Note that for $ p = 1 $, GADC reduces to ADC with only two Kraus operators. It has also two Kraus operators for $ p = 0 $ which describes an amplification process \cite{KSW-PRA-2020}.

In this section, we consider evolution of the generalized $\W\Wb$-state described by
\begin{eqnarray} \label{eq: WWb-state}
| \W\Wb \ra = \cos \theta ~ | \W \ra + \sin \theta ~ e^{i\phi} | \Wb \ra 
\end{eqnarray}
under GADC channel \eqref{eq: GADC-channel} when all qubits are affected by the environment. In this way, 64 Kraus operators describe the effect of this channel on our three-qubit system. Here, $ | \W \ra $ is given by \eqref{eq: W-state} and
\begin{eqnarray} \label{eq: Wb-state} 
| \Wb \ra &=& \frac{1}{\sqrt{3}} ( | 110 \ra + | 101 \ra + | 011 \ra ) 
\end{eqnarray}
is its obverse \cite{DaDoDoAr-PRA-2015}. 
The entanglement properties of the state \eqref{eq: WWb-state} with $ \theta = \pi/4 $ i.e., the permutation symmetric superposition of the W state and its obverse $\Wb$, has already been studied \cite{SuDeRa-PRA-2012, DeSuRa-QIP-2012}. It has been argued that this state belongs to the GHZ entanglement class. Furthermore, it has been used to study the dynamics of quantum coherence and monogamy in a non-Markovian environment \cite{RaChJaByAl-SR-2019}.

For the special values $ \theta = 0 $ and $ \theta = \frac{\pi}{2} $ the generalized $\W\Wb$-state reduces to W-state and $\Wb$-state respectively. One finds all reduced states $ \rho'_{AB} $, $ \rho'_{AC} $ and $ \rho'_{BC} $ are the same being X-shaped. The corresponding concurrence for special values of $ \theta $ recasts
\begin{eqnarray} 
\mathcal{C}(\rho'_{AB}) \bigg |_{\theta = 0 } &=& \max \left\{0, \frac{2}{3} \left( 1 - d - \sqrt{ f_{\W}(d, p) } \right) \right\} , \label{eq: concurAB_W_GADC}
 \\
\mathcal{C}(\rho'_{AB}) \bigg |_{\theta = \pi / 2 } &=& \max \left\{0, \frac{2}{3} \left( 1 - d - \sqrt{ f_{\Wb}(d, p) } \right) \right\} , \label{eq: concurAB_Wb_GADC}
 \\
\mathcal{C}(\rho'_{AB}) \bigg |_{\theta = \pi / 4 } &=& \max \left\{0, \frac{1}{3} \left( 2 - 2 d - \sqrt{ f_{\W\Wb}(d, p) } \right)  \right\} , \label{eq: concurAB_WWb_GADC}
\end{eqnarray}
where, 
\begin{eqnarray}
f_{\W}(d, p) &=&  d (p-1) \left( d^3 (p-1) (1-3 p)^2+2 d^2 p (3 p-1)+d (3-5 p)-2 \right) , \\
f_{\Wb}(d, p) &=&  d p \left(d^3 p (2-3 p)^2-2 d^2 \left(3 p^2-5 p+2\right)+d (2-5 p)+2\right) , \\
f_{\W\Wb}(d, p) &=&  d^4 \left(6 p^2-6 p+1\right)^2+2 d^3 \left(6 p^2-6 p+1\right)-6 d^2 (1-2 p)^2+2 d+1 .
\end{eqnarray}
One notes that function $ f_{\W\Wb}(d, p)$ and thus the concurrence for $ \theta = \pi / 4 $ are symmetric functions of $ p - 1/2 $ for a given $d$. 

The generalized W$\Wb$-state is mapped onto a matrix with rank 8 under GADC. So, there is no known analytical method for computing $ \mathcal{C}^2_{A|BC} $. But, to have an idea we examine this quantity for the {\it spectral decomposition} of the mapped state,
\begin{eqnarray} \label{eq: rhopWWb-sd}
\varrho' &=& \sum_{i=1}^8 P_i | \lam_i \ra \la \lam_i |, 
\end{eqnarray}
$P_i$ being the i$^{\text{th}}$ eigenvalue of the mapped state $\rho'$ and $ | \lam_i \ra $ the corresponding eigenvector. Then, one has that 
\begin{eqnarray} \label{eq: con2AD_rhopWWb-sd}
\mathcal{C}^2_{A|BC}(\varrho') &=& \sum_{i=1}^8 P_i ~ \mathcal{C}^2_{A|BC} (| \lam_i \ra \la \lam_i |) .
\end{eqnarray}
For the spectral decomposition of W-state and $\Wb$-state under GADC one obtains
\begin{eqnarray} 
\mathcal{C}^2_{A|BC}(\varrho'_{\W}) &=& 
\frac{1}{729} \big(90 d^6 (p-1)^4 p^2+150 d^5 (p-1)^3 p - 10 d^4 (p-1)^2 \left(15 p^2-15 p-4\right) 
\nonumber \\
& \qquad &
-d^3 (p-1)^2 (1053 p+80)+d^2 \left(-1742 p^2+1297 p+445\right)
\nonumber \\
& \qquad &
+81 d (5 p-13)+648 \big) , \label{eq: con2AD_rhopW-sd}
\\
\mathcal{C}^2_{A|BC}(\varrho'_{\Wb}) &=& 
\frac{1}{324} \big( 45 d^6 (p-1)^2 p^4+30 d^5 (p-1) p^3+5 d^4 p^2 \left(-6 p^2+6 p+1\right)
\nonumber \\
& \qquad &
+2 d^3 p^2 (252 p-257)+d^2 p (696-787 p)+96 d (p-3) + 288 \big) , \label{eq: con2AD_rhopWb-sd}
\end{eqnarray}
showing that $ \mathcal{C}^2_{A|BC}(\varrho'_{\W})\big|_{p=1} = \mathcal{C}^2_{A|BC}(\varrho'_{\Wb})\big|_{p=0} $.

\section{Results and discussion} \label{sec: results}

\begin{figure} 
\centering
\includegraphics[width=12cm,angle=-0]{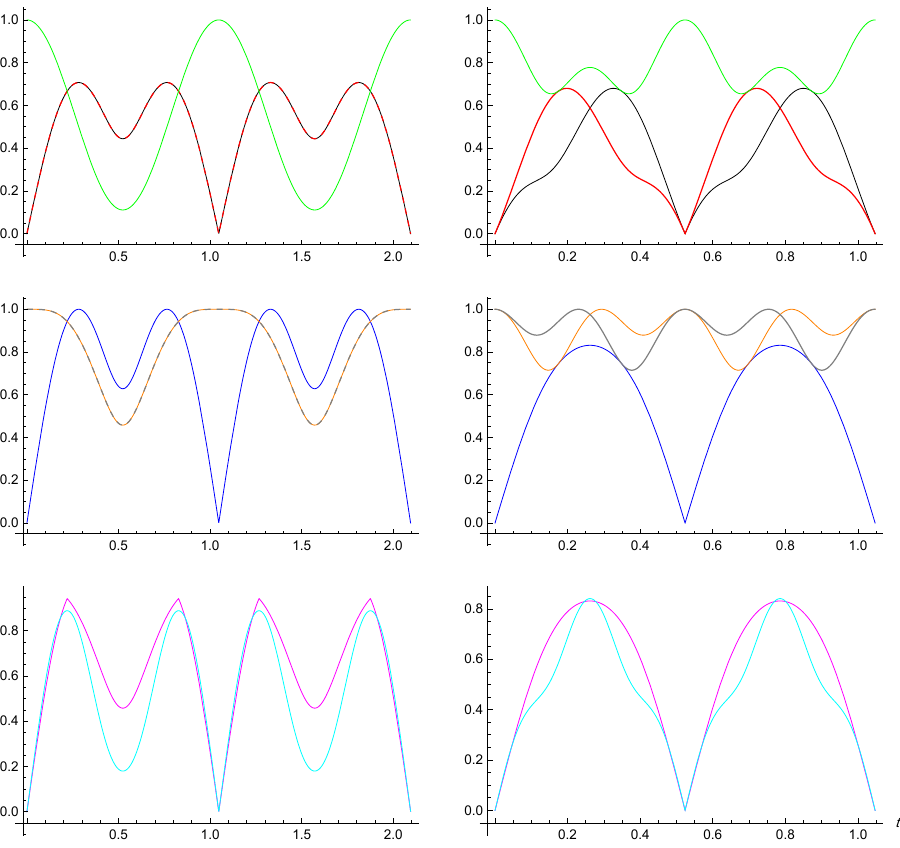}
\caption{
Evolution of concurrences of reduced states (first row), one-to-other concurrences (middle row) and tripartite measures of entanglement (last row) in two periods of oscillation under the Schr\"odinger equation for $J=1$ for $D=0$ (left column) and for $D =\sqrt{3} $ (right column) when the system is initially described by the gW state \eqref{eq: geW-state} with $ a = b = 1 /\sqrt2 $ i.e., $ | 0 \ra \otimes ( | 0 1 \ra + | 1 0 \ra ) / \sqrt{2} $. 
Color codes are as follows: $ \mathcal{C}_{AB}(t) $ (black), $ \mathcal{C}_{AC}(t) $ (red), $ \mathcal{C}_{BC}(t) $ (green), $ \mathcal{C}_{A|BC}(t) $ (blue), $ \mathcal{C}_{B|AC}(t) $ (orange), $ \mathcal{C}_{C|AB}(t) $ (gray), the genuine tripartite concurrence $ \mathcal{C}_{\GTC}(t) $ (magenta) and the concurrence fill $ \mathcal{F}(t) $ (cyan).
}
\label{fig: pSch} 
\end{figure}

In this section, numerical results are presented for various entanglement measures for $ \hb = 1 $.
In figure \ref{fig: pSch} we have studied evolution of different measures of entanglement under the Schr\"odinger equation with the interaction Hamiltonian \eqref{eq: Ham}. System is initially described by the gW state \eqref{eq: geW-state} with $ a = b = 1 /\sqrt2 $ i.e.,
\begin{eqnarray} \label{eq: gW_a=b}
|\psi (0) \ra &=& \frac{1}{\sqrt{2}} | 0 \ra \otimes ( | 0 1 \ra + | 1 0 \ra ). 
\end{eqnarray}
As we analysed in Section \ref{sec: en_ev_Sch}, entanglement measures are all independent of both the anisotropy parameter and the magnetic field strength; and are periodic in time. Furthermore, the three-tangle is zero.
From \eqref{eq: gW_a=b} one can easily see that initial reduced states $ \rho_{AB}(0) $ and $ \rho_{AC}(0) $ are the same being disentangled while the initial reduced state $ \rho_{BC}(0) $ is a Bell state with maximal entanglement. This point is clearly seen in the figure. In the absence of DM interaction, reduced states $ \rho_{AB}(t) $ and $ \rho_{AC}(t) $ remain the same at all times, while DM interaction breaks this symmetry.
Eq. \eqref{eq: gW_a=b} reveals that the one-to-other concurrence $ \mathcal{C}_{A|BC} $ is initially zero as one expects since the first qubit is initially disentangled from the two others. 
Furthermore, as the middle panel of the left column displays, one has $ \mathcal{C}_{B|AC}(t) = \mathcal{C}_{C|AB}(t) $ for $D=0$.
Note that the break in the curve of the genuine tripartite concurrence $ \mathcal{C}_{\GTC}(t) $, the magenta curve of the left column of the figure, is due to the minimization scheme that needs to be done according to the definition \eqref{eq: C-gtc}.
For our parameters, when the DM interaction is in place, the one-to-other concurrence $ \mathcal{C}_{A|BC}(t) $ is always equal or less than the two other one-to-other concurrences.
Therefore, it can be inferred that the minimum argument in the expression for GTC does not result in any breaks in its plot. In fact, one has that  $ \mathcal{C}_{\GTC}(t) = \mathcal{C}_{A|BC}(t) $. 
The initial value of the concurrence fill is zero, as one edge of the concurrence triangle has a zero length, resulting in a zero area of the concurrence triangle.
Comparison of the left and right panels reveals that the period of entanglement measures becomes shorter when the DM interaction is in place. Furthermore, this interaction affects the maximal value of entanglement and also its time of occurrence; $ \frac{2\sqrt{2}}{3} \approx 0.9428 $ ($ \frac{2\sqrt{14}}{9} \approx 0.8315 $) and $ \frac{8}{9} \approx 0.8889 $ ($ \frac{ 28 \sqrt[4]{35} }{81} \approx 0.8408 $) respectively for GTC and concurrence fill, at times 0.2197 and 0.2618 without (with) DM interaction.

%==========================================
%============================================
%

\begin{figure} 
\centering
\includegraphics[width=12cm,angle=-0]{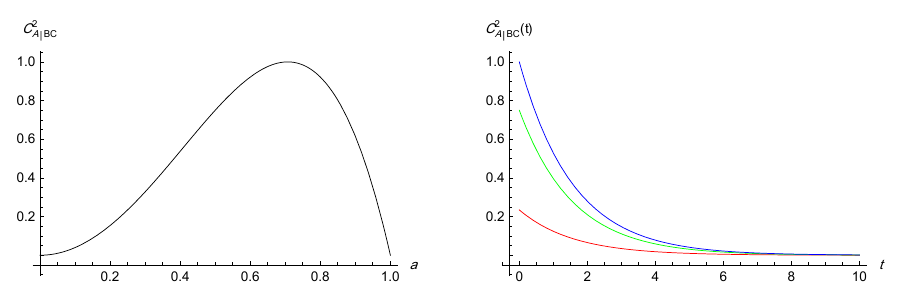}
\caption{
The squared concurrence between parties $A$ and $BC$, as a single object, in the framework of the von Neumann formalism (left panel) and the Milburn formalism (right panel) when the system is initially described by the gGHZ state \eqref{eq: geGHZ-state}.
Left panel: \eqref{eq: C2_A|BC_rhot_vN} in terms of the superposition coefficient $ a $.
Right panel: Evolution of \eqref{eq: C2_A|BC_rhot} for $ \ga = 0.5 $ and $ B = 0.1 $ for $ a = 0.25 $ (red curve), $ a = 0.5 $ (green curve) and $ a = 1 / \sqrt{2} $ (blue curve). The only relevant parameter of the Hamiltonian is the magnetic field strength $B$. 
}
\label{fig: pCGHZMil} 
\end{figure}

In figure \ref{fig: pCGHZMil} we have examined $I$-tangle $  \mathcal{C}_{A|BC}^2 $ within the framework of the von Neumann (left panel) and the Milburn (right panel). The system is initially described by the gGHZ state \eqref{eq: geGHZ-state}.
The state reduces to $ | 111 \ra $ and  $ | 000 \ra $ respectively for $a=0$ and $a=1$ implying no entanglement between subsystems. The $I$-tangle takes its maximum value one for %$ a = 1 / \sqrt{2} $ i.e., when the system is described by the conventional 
the GHZ state.
Left panel depicts variation of $I$-tangle with the superposition coefficient $ a $ appearing in \eqref{eq: geGHZ-state}. 
Note that within the standard quantum mechanics, $I$-tangle is independent of time. 
Right panel depicts evolution of the $I$-tangle %for the strength magnetic field $ B = 0.1 $ 
under the Milburn equation for $ B = 0.1 $ and $ \ga = 0.5 $ but for different values of the superposition coefficient $ a $. The figure illustrates that the squared one-to-other concurrence decreases with time for a given value of $a$, as expected from a decoherence structure and is maximal for the gGHZ state.

% 
%==========================================
%============================================

%The left panel of Figure \ref{fig: pCGHZMil2} shows the evolution of the $I$-tangle for a given magnetic field intensity but for different values of the intrinsic decoherence parameter, while the right panel considers the opposite scenario, where the intrinsic decoherence parameter is constant and the magnetic field varies.

\begin{figure} 
\centering
\includegraphics[width=12cm,angle=-0]{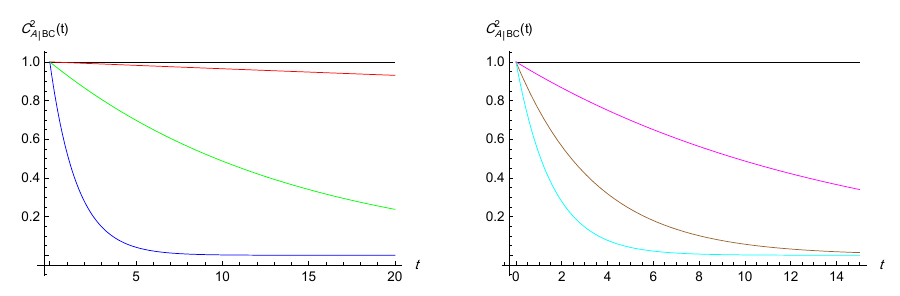}
\caption{
The squared one-to-other concurrence \eqref{eq: C2_A|BC_rhot} under the intrinsic decoherence when the system is initially described by the GHZ state i.e., \eqref{eq: geGHZ-state} with $ a = 1 / \sqrt{2} $. In the left panel intensity of the applied magnetic field is fixed to $ B = 0.1 $,  while in the right one the intrinsic decoherence parameter is constant, $ \ga = 0.5 $.  
Left panel: $ \ga = 100 $ (red curve), $ \ga = 5 $ (green curve) and $ \ga = 0.5 $ (blue curve).  
Right panel: $ B = 0.1 $ (magenta curve), $ B = 0.2 $ (brown curve) and $ B = 0.3 $ (cyan curve). 
In both panels the black curve corresponds to the standard quantum mechanical result.
}
\label{fig: pCGHZMil2} 
\end{figure}

The left panel of Figure \ref{fig: pCGHZMil2} shows the evolution of the $I$-tangle for a given magnetic field intensity but for different values of the intrinsic decoherence parameter. In contrast, the right panel examines the reverse situation, where the intrinsic decoherence parameter remains constant while the magnetic field changes.
The tripartite system is initially described by the GHZ state. 
Again, the squared one-to-other concurrence decreases with time in both cases. Furthermore, in a given $t$ and given $B$ ($\ga^{-1}$) the $I$-tangle decreases with $\ga^{-1}$ ($B$). Additionally, at any given instant of time, the rate of decrement grows with $\ga^{-1}$ ($B$) in a given $B$ ($\ga^{-1}$).

%==========================================
%============================================

\begin{figure} 
\centering
\includegraphics[width=12cm,angle=-0]{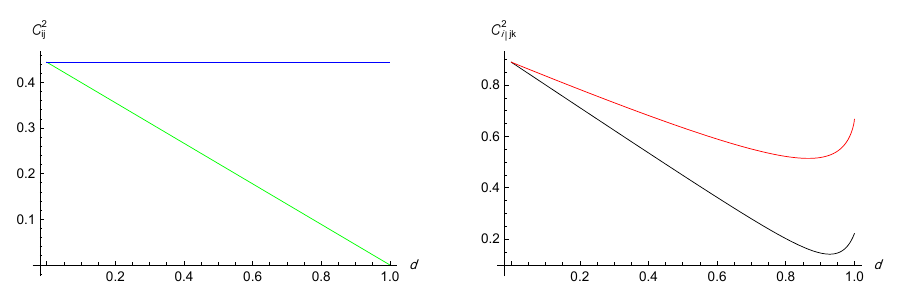}
\caption{
The squared concurrences of reduced states (left) and the squared one-to-other concurrences (right) in terms of the decoherence strength $d$ when the system is initially described by the $\W$ state \eqref{eq: W-state} and undergo PDC-I given by \eqref{eq: channel-I}.
Color codes are as follows;
green: $ \mathcal{C}^2(\rho'_{AB}) $ given by \eqref{eq: C_AB-rhopW}, blue: $ \mathcal{C}^2(\rho'_{BC}) $ given by \eqref{eq: C_BC-rhopW}, black: $ \mathcal{C}^2_{A|BC}(\rho'_{\W}) $ given by \eqref{eq: C2_A|BC-rhopW} and red: $ \mathcal{C}^2_{B|AC}(\rho'_{\W}) $ given by \eqref{eq: C2_B|AC-rhopW}.
}
\label{fig: rhoW-PDCI} 
\end{figure}

Figure \ref{fig: rhoW-PDCI} examines concurrences under PDC-I where only the first qubit is affected by the phase damping channel. The system is initially described by the $\W$ state \eqref{eq: W-state}. In the left panel we have plotted squared reduced concurrences in terms of the decoherence strength $d$ while right panel displays one-to-other concurrences versus the same parameter. As one expects the entanglement corresponding to the reduced state $ \rho'_{BC} $ represented by the blue curve in the right panel is not affected by the channel since only the first qubit interacts with the environment. The quantity $ \mathcal{C}^2(\rho'_{AB}) $ decreases with increasing $d$ and reaches zero when $d=1$. On the other hand, as the right panel displays, all squared one-to-other concurrences represent a minimum before the decoherence strength reaches its maximum value. The minimal values and their positions are as follows:
\begin{equation*}
\mathcal{C}^2_{A|BC} \bigg|_{\min} \approx 0.1420 ~\text{for}~ d \approx 0.9280, \qquad
\mathcal{C}^2_{B|AC} \bigg|_{\min} \approx 0.5146 ~\text{for}~ d \approx 0.8656 .
\end{equation*}
An interesting phenomenon occurs here: in a certain interval of the decoherence strength $d$, the environment has constructive effect on the one-to-other entanglements. 
Unlike the reduced entanglements, the one-to-other entanglements never vanish.

%==========================================
%============================================
%

\begin{figure} 
\centering
\includegraphics[width=12cm,angle=-0]{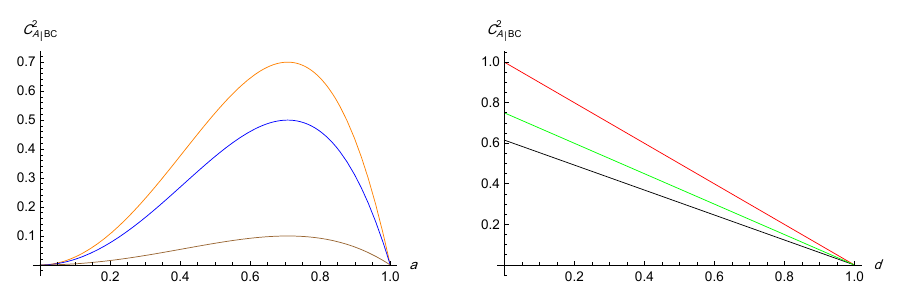}
\caption{
The squared one-to-other concurrence \eqref{eq: C2_A|BC-rhopGHZ} for the gGHZ state under ADC-I versus the superposition coefficient $a$ (left panel) and the decoherence strength $d$ (right panel) for $ d = 0.3 $ (orange), $ d = 0.5 $ (blue) and $ d = 0.9 $ (brown); and for $ a = 0.9 $ (black), $ a = \frac{1}{\sqrt{2}} \approx 0.71 $ (red) and $ a = 0.5 $ (green).
}
\label{fig: rhoGHZ-ADC} 
\end{figure}

\begin{figure} 
\centering
\includegraphics[width=12cm,angle=-0]{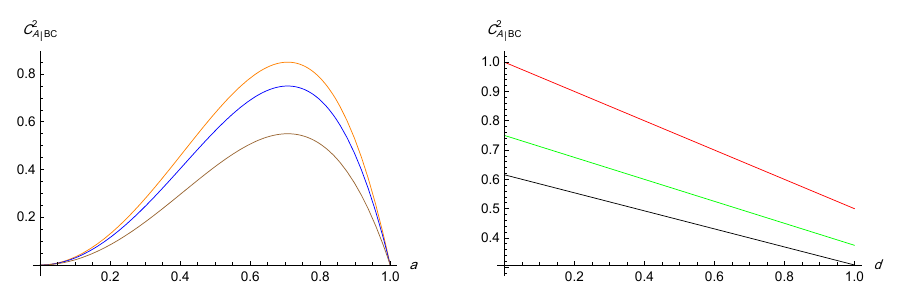}
\caption{
The squared one-to-other concurrence \eqref{eq: C2_A|BC-rhopGHZ-3} for the gGHZ state under ADC-III \eqref{eq: channel-III} versus the superposition coefficient $a$ (left panel) for $ d = 0.3 $ (orange), $ d = 0.5 $ (blue) and $ d = 0.9 $ (brown); and the decoherence strength $d$ (right panel) for $ a = 0.9 $ (black), $ a = \frac{1}{\sqrt{2}} $ (red) and $ a = 0.5 $ (green).
}
\label{fig: rhoGHZ-ADC-3} 
\end{figure}

Figures \ref{fig: rhoGHZ-ADC} and \ref{fig: rhoGHZ-ADC-3} examine the behaviour of 
%the squared one-to-other concurrence 
$ C^2_{A|BC} $ in terms of the expansion coefficient $a$ (left panels) and the decoherence strength $d$ (right panels) for the gGHZ state \eqref{eq: geGHZ-state} respectively under the effect of ADC-I and ADC-III.
For a given $d$, the curve of $ \mathcal{C}^2_{A|BC} $ in terms of $a$ represents a maximum at $ a = 1 /\sqrt{2} $, where the maximal value decreases with $d$.
This quantity is a decreasing function of the decoherence strength $d$ for a given $a$.
Furthermore, from Eqs. \eqref{eq: C2_A|BC-rhopGHZ} and \eqref{eq: C2_A|BC-rhopGHZ-3} one has that
\begin{eqnarray*}
\mathcal{C}^2_{A|BC} \bigg|_{ \text{ADC-I} } &=& 2 \frac{1-d}{2-d} ~ \mathcal{C}^2_{A|BC} \bigg|_{ \text{ADC-III} } ,
\end{eqnarray*}
showing that for a given $d$, ADC-I affects the system more strongly than ADC-III. This indicates that the damping rate is greater when the first qubit is affected by the environment than the last one.
It is worth mentioning that for both channels all bipartite reduced states are X-shaped with zero concurrence. Therefore, the above result shows that the residual entanglement is higher for ADC-III in comparison to ADC-I.
This examination suggests that, when the focus qubit is the only one impacted by the channel, the influence of the environment is stronger.

%==========================================
%============================================
% 
%==========================================
%============================================
%

\begin{figure} 
\centering
\includegraphics[width=12cm,angle=-0]{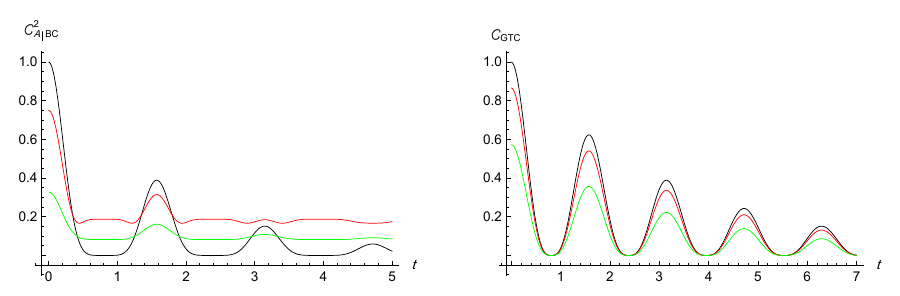}
\caption{
Evolution of the squared concurrence \eqref{eq: C2_A|BC-nonMarkov_gGHZ} (left panel) and the genuine tripartite concurrence \eqref{eq: gen_con_rhoGHZg_nonMarkov} (right panel) for the gGHZ state under the non-Markovian dephasing channel \eqref{eq: nonMarkov-ch} for $ b = 1 $ and $ \tau = 5 $ and for different values of the superposition coefficient, $a = \frac{1}{\sqrt{2}} $ (black), $ a = 0.5 $ (red) and $ a = 0.3 $ (green).
}
\label{fig: p_nonMarkov_gGHZ} 
\end{figure}

We have examined in Fig. \ref{fig: p_nonMarkov_gGHZ}, evolution of the squared one-to-other concurrence \eqref{eq: C2_A|BC-nonMarkov_gGHZ} (left panel) and the genuine tripartite concurrence \eqref{eq: gen_con_rhoGHZg_nonMarkov} (right panel) under the non-Markovian dephasing channel \eqref{eq: nonMarkov-ch} for different values of the expansion coefficient $a$ when the system is initially described by the gGHZ state \eqref{eq: geGHZ-state}.
For $a=1/\sqrt{2}$, %i.e., when the system is described by the GHZ state, 
$ \mathcal{C}^2_{A|BC}$ represents dark periods \cite{FiTa-PRA-2006}. However, this is not the entanglement {\it sudden} death phenomenon since there is no {\it break} in the curve. 
Revivals of entanglement after these finite dark periods is mainly due to the memory effect of the non-Markovian environment
\cite{Pa-QIP-2016, BeFrCo-PRL-2007}.
In a given instant of time, GTC is maximal for the GHZ state.
Asymptotic values of the squared one-to-other concurrence $ \mathcal{C}^2_{A|BC} $ is $0$, $ 0.1875 $ and $ 0.0819 $ respectively for $ a = 1 /\sqrt{2} $, $ a = 0.5 $ and $ a = 0.3 $. For all values of $a$, GTC goes to zero at the long time limit. 

% 
%==========================================
%============================================

% 
%Figure \ref{fig: p_nonMarkov_GHZ_mix} demonstrates evolution of the squared one-to-other concurrence \eqref{eq: C2A_BC_nonMar_GHZ_mix} corresponding to the mixture \eqref{eq: GHZ_000_mix} under the non-Markovian dephasing channel \eqref{eq: nonMarkov-ch} for two different inverse fluctuation rates: $ \tau = 2 $ (left panel) and $ \tau = 20 $ (right panel); and for different values of the weight factor appearing in the mixture.

\begin{figure} 
\centering
\includegraphics[width=12cm,angle=-0]{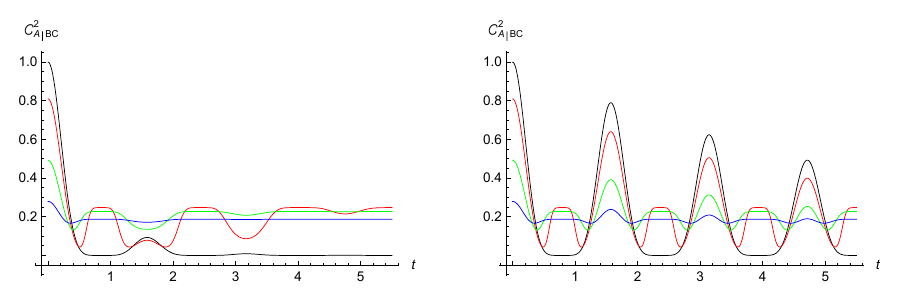}
\caption{
Evolution of the squared one-to-other concurrence \eqref{eq: C2A_BC_nonMar_GHZ_mix} corresponding to the mixture \eqref{eq: GHZ_000_mix} under the non-Markovian dephasing channel \eqref{eq: nonMarkov-ch} for channel parameters $ \tau = 2 $ (left panel), $ \tau = 20 $ (right panel) and $ b = 1 $ for different values of the weight factor, $ w = 0 $ (black), $ w = 0.1 $ (red), $ w = 0.3 $ (green) and $ w = 0.5 $ (blue).
}
\label{fig: p_nonMarkov_GHZ_mix} 
\end{figure}

Figure \ref{fig: p_nonMarkov_GHZ_mix} illustrates the evolution of the squared one-to-other concurrence \eqref{eq: C2A_BC_nonMar_GHZ_mix} associated with the mixture \eqref{eq: GHZ_000_mix} under the non-Markovian dephasing channel \eqref{eq: nonMarkov-ch}. This is shown for two distinct inverse fluctuation rates: $ \tau = 2 $ (left panel) and $ \tau = 20 $ (right panel), along with varying values of the weight factor present in the mixture.
%From the expression of $\Lam$ in terms of $\uptau$ one sees that increasing this parameter makes $\Lam$ to 
Dark periods of entanglement is seen only for the GHZ state and interestingly enough steady state concurrence is non-zero for the mixture while vanishes when the system is initially described by the pure GHZ state.
Decreasing the the fluctuation rate $\tau^{-1}$ increases the number of oscillations and also delays reaching the steady state. The steady state value of entanglement decreases with $w$ which is acceptable since increasing $w$ means the weight factor of the fully separable state $|000\ra$ enhances while that of the entangled GHZ state diminishes. 
%
%==========================================
%============================================

%

\begin{figure} 
\centering
\includegraphics[width=12cm,angle=-0]{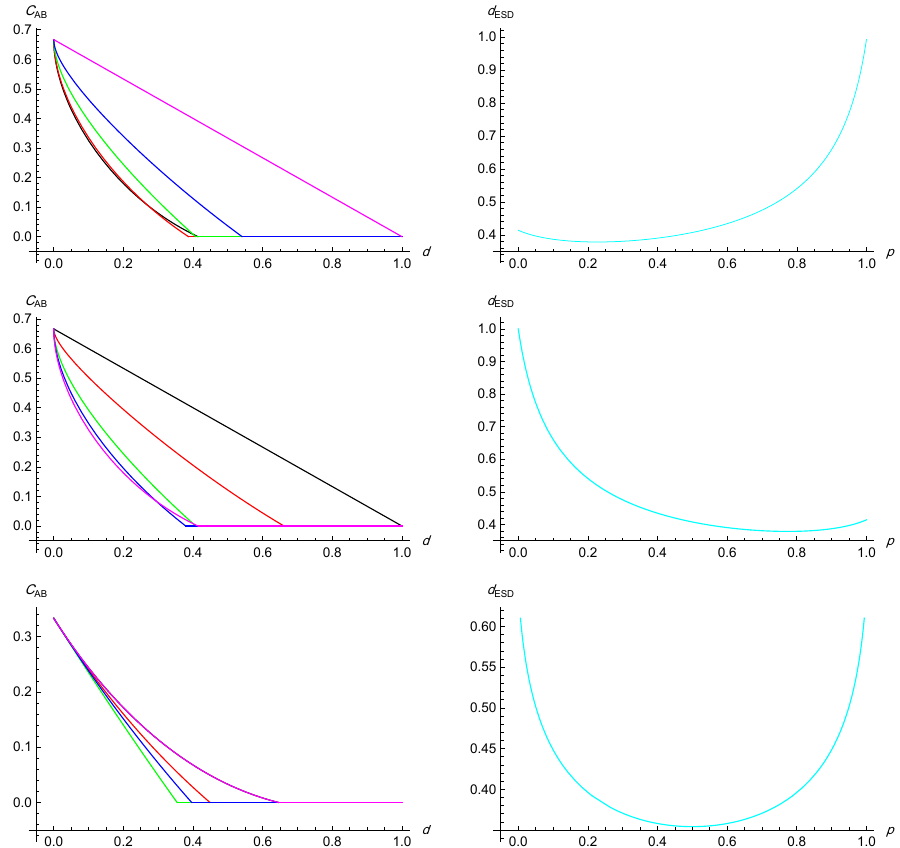}
\caption{
Reduced concurrence $ \mathcal{C}_{AB} $ (left panels) and the decoherence strength at ESD point (right panels) under GADC when the system is initially described by the generalized $\W\Wb$-state \eqref{eq: WWb-state} for $ \theta = 0 $ (top panels), $ \theta = \frac{\pi}{2} $ (middle panels) and $ \theta = \frac{\pi}{4} $ (bottom panels) for $ p = 0 $ (black curves), $ p = 0.1 $ (red curves), $ p = 0.5 $ (green curves), $ p = 0.8 $ (blue curves) and $ p = 1 $ (magenta curves).
}
\label{fig: pGADCall} 
\end{figure}

\begin{figure} 
\centering
\includegraphics[width=12cm,angle=-0]{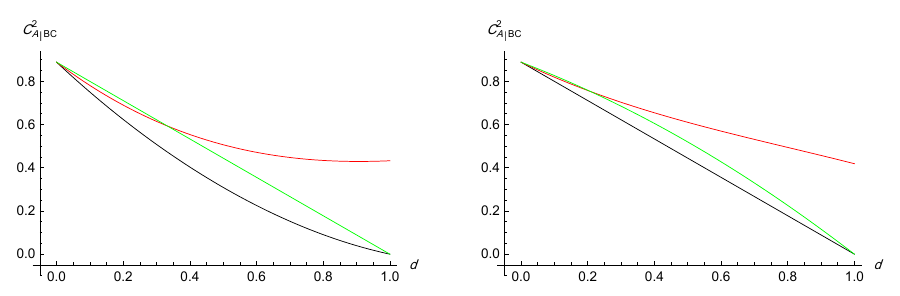}
\caption{
Evolution of the squared on-to-other concurrences \eqref{eq: con2AD_rhopW-sd} and \eqref{eq: con2AD_rhopWb-sd} respectively for the W-state (left panel) and $\Wb$-state (right panel) under GADC for $ p = 0 $ (black), $ p = 0.5 $ (red) and $ p = 1 $ (green).
}
\label{fig: pcon2AD_WWb_sd} 
\end{figure}

Finally, in Figures \ref{fig: pGADCall} and \ref{fig: pcon2AD_WWb_sd}, we have examined a system consisting of three qubits, initially described by the generalized $\W\Wb$-state \eqref{eq: WWb-state}, under the influence of the generalized amplitude damping channel \eqref{eq: GADC-channel}.
Left panels of Fig. \ref{fig: pGADCall} depict the reduced concurrence $ \mathcal{C}_{AB} $ in terms of the decoherence strength $d$ for different values of the channel parameter $p$ for W state (top panel), its obverse state $\Wb$ (middle panel) and the permutation symmetric superposition of them (bottom panel). The phenomenon of ESD is clearly seen for certain values of $p$. Right panels of this figure display the position of ESD in terms of $p$ for different initial states.
Minimum value of $ d_{\ESD} $ is $ 0.3787 $ for both W and $\Wb$ states; and $ 0.3542 $ for $ \theta = \pi /4 $, taking place respectively at $ p = 0.2265 $, $ p = 0.7734 $ and $ p = 0.5 $.
A comparison with Ref. \cite{LiShZo-PLA-2010} effectively underscores the novelty of our findings. This study examines BNSD in the context of intrinsic decoherence and shows that violations of Bell inequalities are completely suppressed within a specific timescale, which is affected by both intrinsic decoherence and the external magnetic field. In contrast, our research reveals that ESD for reduced concurrences occurs under the GADC when the system is initialized in the generalized $\W\Wb$-state \eqref{eq: WWb-state}. While the suppression of BNSD in the previous study is determined by intrinsic decoherence, the occurrence of ESD in our work is dictated by the parameters of the GADC.
Figure \ref{fig: pcon2AD_WWb_sd} studies evolution of the squared one-to-other concurrences \eqref{eq: con2AD_rhopW-sd} and \eqref{eq: con2AD_rhopWb-sd} respectively for $W$ and $\Wb$ states for different values of $p$. Note that %none of these states are mapped to a state of a rank smaller or equal to two under GADC. In fact,% 
 we have plotted the one-to-other concurrence corresponding to the spectrally decomposed mapped state \eqref{eq: rhopWWb-sd}.
For $p=0$ and $p=1$ that GADC reduces to a channel with only two Kraus operators corresponding respectively to an amplification process \cite{KSW-PRA-2020} and the amplitude damping channel, $I$-tangle vanishes for $d=1$ for both $W$ and $\Wb$ states.
%
%==========================================
%============================================
%
%==========================================
%============================================

\section{Summary and conclusions} \label{sec: conclusion}

This work investigates a tripartite system of three qubits initialized in either a gGHZ state, a W state, or a mixture of these with fully separable states. The system's evolution is analyzed using the Schr\"odinger equation with an XXZ Heisenberg chain Hamiltonian that includes DM interaction. It is found that the gGHZ state's entanglement remains static, while the gW state's entanglement exhibits periodic behavior influenced by coupling strength and DM interaction. Other Hamiltonian parameters do not impact the entanglement measures. The study also examines the effects of intrinsic decoherence on the gGHZ state and its mixtures, revealing that only magnetic field strength influences entanglement measures, with decoherence suppressing entanglement in a rate-dependent manner. Additionally, different decoherence channels are explored, showing that one-to-other concurrences generally decrease with decoherence strength, except for the W state under PDCI, which shows a minimum. The GHZ state under non-Markovian dephasing exhibits finite dark periods, while steady-state concurrence remains for mixtures but vanishes for the pure GHZ state. Lastly, the research considers a generalized $\W\Wb$-state under generalized amplitude damping, resulting in a density matrix of rank eight. Squared one-to-other concurrences for the spectral decomposed state and reduced concurrences are calculated, revealing the phenomenon of ESD.
\\
\\
{\bf Acknowledgements:} The author is grateful to reviewers for their constructive comments, useful suggestions, and pointing to important references. Support from the University of Qom is acknowledged. 
\\
\\
{\bf Data availability}: This manuscript has no associated data.

%%===========================================================================================%%

\begin{appendices}

%\section{ Entanglement measure for rank-2 mixed states }

\section{ Matrix $M$ for the generalized GHZ state under the intrinsic decoherence } \label{app: M-matrix}

Consider $ \rho $ to be a state of a three-qubit system having at most two nonzero eigenvalues, a second rank matrix, 
\begin{eqnarray}
\rho &=& \lam_1 | \lam_1 \ra \la \lam_1 | + \lam_2 | \lam_2 \ra \la \lam_2 | .
\end{eqnarray}
As discussed in \cite{Os-PRA-2005}, one can compute the squared one-to-other concurrence $ \mathcal{C}^2_{A|BC} $ as follows. First, construct the tensor
\begin{eqnarray} \label{eq: tenT-0}
T_{ijkl} &=& \tr( \ga_{ij} \ti{\ga}_{kl} ),
\end{eqnarray}
where $ \ga_{ij} = | \lam_i \ra \la \lam_j |  $ is the outer product of the eigenvectors of $ \rho $ and
\begin{eqnarray} \label{eq: ga-tilde}
\ti{\ga}_{ij} &=& \tr( \ga_{ij}^{\dag} ) \mathds{1}_2 \otimes \mathds{1}_4 - \tr_{BC}( \ga_{ij}^{\dag} ) \otimes \mathds{1}_4 - \mathds{1}_2 \otimes \tr_A( \ga_{ij}^{\dag} ) + \ga_{ij}^{\dag} .
\end{eqnarray}
Using \eqref{eq: ga-tilde} in \eqref{eq: tenT-0} yields
\begin{eqnarray} \label{eq: tenT}
T_{ijkl} &=& \tr( \ga_{ij} ) \tr( \ga_{kl}^{\dag} ) - \tr_{BC}[ \tr_A( \ga_{ij} ) \tr_A( \ga_{kl}^{\dag} ) ] - \tr_A [ \tr_{BC}( \ga_{ij} ) \tr_{BC}( \ga_{kl}^{\dag} ) ] + \tr( \ga_{ij} \ga_{kl}^{\dag} ), \nonumber \\
\end{eqnarray}
where $ \tr_A( \ga_{ij} ) $ and $ \tr_{BC}( \ga_{ij} ) $ display partial traces of $ \ga_{ij} $ over respectively the subsystems $A$ and $BC$, as a single object.
%
%Osborne \cite{Os-PRA-2005} has proved that the square of concurrence between parties $A$ and $BC$, with the last treated as a single system is given by
%
%\begin{eqnarray} \label{eq: C2_A|BC_rank2}
%C^2_{A|BC} &=& \tr( \rho \ti{\rho} ) + 2 m_{\min} [ 1 - \tr( \rho^2 ) ]
%\end{eqnarray}
%
%where
%
%\begin{eqnarray}
%\ti{\rho} &=& \mathds{1}_2 \otimes \mathds{1}_4 - \rho_A \otimes \mathds{1}_4 - \mathds{1}_2 \otimes \rho_{BC} + \ti{\rho},
%\end{eqnarray}
%
%$ \rho_A $ and $ \rho_{BC} $ being reduced states describing respectively parties $ A $ and $ BC $; and $  m_{\min} $ is the minimum eigenvalue of the real symmetric matrix 
%
From the tensor \eqref{eq: tenT} one then constructs the real symmetric matrix
\begin{eqnarray}
M_{11} &=& \frac{1}{4} ( T_{1221} + 2 T_{1122} + T_{2112} ),  \label{eq: M11}  \\
M_{12} &=& \frac{i}{4} ( T_{1221} - T_{2112} ),  \label{eq: M12} \\
M_{13} &=& \frac{1}{4} ( T_{1121} - T_{2122} + T_{1112} - T_{1222} ),  \label{eq: M13}  \\
M_{22} &=& - \frac{1}{4} ( T_{1221} - 2 T_{1122} + T_{2112} ),   \label{eq: M22} \\
M_{23} &=& \frac{i}{4} ( T_{1121} - T_{1112} + T_{2122} - T_{1222} ), \label{eq: M23} \\
M_{33} &=& \frac{1}{4} ( T_{1111} - T_{1122} + T_{2222} ),  \label{eq: M33} 
\end{eqnarray}
where according to \eqref{eq: C2_A|BC_rank2}, its minimum eigenvalue is used for computation of $I$-tangle.

%\subsection{ Matrix $M$ for GHZ state under intrinsic decoherence }

The independent elements of the matrix $M$ for the generalized GHZ state under the intrinsic decoherence, \eqref{eq: rhot}, recast
\begin{numcases}~
M_{11}  = \frac{ (1-2a^2)^2 - 4a^2(1-a^2) |f_{18}(t)|^2 }{ 2(1-2a^2)^2 +8a^2(1-a^2) |f_{18}(t)|^2 },
 \label{eq: M11_rhot}
\\ 
M_{12}  = 0  \label{eq: M12_rhot}
\\
M_{13}  = \frac{ 2 a \sqrt{ 1 - a^2 } (1-2a^2) |f_{18}(t)| }{ (1-2a^2)^2 + 4 a^2(1-a^2) |f_{18}(t)|^2 },
\label{eq: M13_rhot}
 \\
M_{22}  = \frac{1}{2}, \label{eq: M22_rhot}
\\
M_{23} = 0, \label{eq: M23_rhot} 
\\
M_{33} = - M_{11}. \label{eq: M33_rhot}
\end{numcases}
The minimum eigenvalue of this matrix is the constant value 
\begin{eqnarray} \label{eq: mmin_rhot}
m_{\min} &=& - \frac{1}{2},
\end{eqnarray}
being independent of time; and parameters of the Hamiltonian and the initial state.

\end{appendices}

\end{document}